\documentclass[%
 reprint,
superscriptaddress,xcolor=dvipsnames,
 amsmath,amssymb,
 aps,
 pra,
]{revtex4-2}

\usepackage{comment}
\usepackage[dvipsnames]{xcolor}
\usepackage{graphicx}
\usepackage{dcolumn}
\usepackage{bm}
\usepackage{braket}

\definecolor{applegreen}{rgb}{0.55, 0.71, 0.0}

\begin{document}


\title{Torus Knot Angular Momentum in Twisted Attosecond Pulses from High Harmonic Generation\\
}
\author{Bj\"orn Minneker}\thanks{bjoern.minneker@uni-jena.de}%
\affiliation{Helmholtz Institut, Jena}
\affiliation{GSI Helmholtzzentrum für Schwerionenforschung GmbH, Darmstadt}
\affiliation{Theoretisch Physikalisches Institut, Friedrich-Schiller-Universit\"at, Jena}

\author{Birger B\"oning}%
\affiliation{Helmholtz Institut, Jena}
\affiliation{GSI Helmholtzzentrum für Schwerionenforschung GmbH, Darmstadt}
\author{Anne Weber}
\affiliation{Theoretisch Physikalisches Institut, Friedrich-Schiller-Universit\"at, Jena}

\author{Stephan Fritzsche}%
\affiliation{Helmholtz Institut, Jena}
\affiliation{GSI Helmholtzzentrum für Schwerionenforschung GmbH, Darmstadt}
\affiliation{Theoretisch Physikalisches Institut, Friedrich-Schiller-Universit\"at, Jena}

\date{\today}

\begin{abstract}
Bicircular twisted Laguerre-Gaussian beams possess a definite torus knot angular momentum (TKAM) as a new form of angular momentum. TKAM is conserved in nonlinear atomic processes such as high harmonic generation and can be classified by a time delay parameter $\tau$ and a coordination parameter $\gamma$. These parameters are defined by the respective projected orbital angular momentum and the energy of the two superimposed Laguerre-Gaussian beams. We derive a consistent geometric method to determine $\tau$ and $\gamma$ from the driving beam as well as from the high harmonic radiation. This method relates both invariance parameters ($\tau$ and $\gamma$) to the emitted high harmonic radiation. Therefore, $\tau$ and $\gamma$ can be read off of two different torus knots. These knots can be constructed from the spatio-temporal evolution of the electric field of the respective high harmonic radiation or the driving beam. We demonstrate the classification of the invariance parameters for a planar atomic gas target irradiated by bicircular Laguerre-Gaussian beams explicitly. Furthermore, we demonstrate that the respective torus knots determined by $\tau$ and $\gamma$ can be mapped onto each other within minor modifications. This geometric method yields a different way to interpret the invariance parameters $\tau$ and $\gamma$ as well as their underlying relation compared to a purely formal derivation. The investigations presented in this work are in good agreement with previous findings and provide insight into the dynamical symmetry of TKAM in the context of high harmonic generation induced by bicircular twisted Laguerre-Gaussian beams. 

\end{abstract}

\pacs{Valid PACS appear here}

\maketitle\section{\label{sec:Introduction}Introduction}
In recent years, strong-field processes like above-threshold ionization \cite{Agostini1979,Paulus1994,Agostini1979} and high harmonic generation (HHG) \cite{McPherson1987,Ferray1988,Frolov2018,Milo2018,Rego2019,Dorney2018} have gained in importance. In particular, high-energetic radiation emitted in the HHG process can nowadays be tailored quite precisely by utilizing the driving beam properties \cite{Paufler2018,Paufler2019,Garcia2017,Gauthier2017}.

It is well known that light beams, as electromagnetic waves, carry energy.
In addition, they possess different forms of momentum \cite{Devlin2017,Ballantine2016}. Besides linear momentum, light beams may carry spin angular momentum (SAM) which is associated with their local polarization properties, as well as orbital angular momentum (OAM). The latter is associated with an azimuthally varying phase of the beam. As a result, such \textit{twisted} light beams to exhibit a phase singularity on the beam axis as well as helical phase fronts \cite{Allen1992,Harris1994,Bazhenov1990}.

These types of angular momenta provide a broad range of variability in strong-field processes like HHG \cite{Paufler2019}.
Especially HHG driven by twisted light beams attracted particular interest in the last years. The work of Pisanty et al. \cite{Pisanty2019a,Pisanty2019} offered a new way to classify high harmonic radiation by a form of conserved angular momentum, namely the torus knot angular momentum (TKAM). TKAM solves crucial classification issues of HHG radiation driven by bicircular bichromatic beams. More precisely, a set of selection rules can be generalized and reduced to only one for the TKAM. In addition, the mathematical description of HHG with bicircular twisted light beams also significantly improves. Moreover, Pisanty et al.\cite{Pisanty2019} demonstrated that TKAM is conserved in HHG.
In order to show this conservation of TKAM, Pisanty et al. explicitly calculated the HHG spectrum and validated the conservation of TKAM in HHG. However, the detailed dynamics which cause the conservation of TKAM remain unknown. Therefore, an intuitive model and a comprehensive discussion concerning TKAM in high harmonic radiation are still missing.

 In this paper, we therefore derive a means to interpret TKAM in high harmonic radiation within a geometric approach. Furthermore, we will in detail discuss the dynamical symmetry of the TKAM in HHG. We derive an intuitive way to interpret TKAM and map the corresponding TKAM invariance parameters $\gamma$ and $\tau$, separately, onto a torus knot. These invariance parameters determine the dynamical symmetry of the driving beam and the high harmonic radiation. Therefore, $\gamma$ and $\tau$ are crucial in the description of TKAM in high harmonic radiation. We explicitly demonstrate the mapping for the high harmonic radiation. However, the mapping can be analogously applied to the driving beam as well. The constructed torus knot determines the eigenvalues of the associated TKAM operator uniquely. To show this, we focus on the classification of high harmonic radiation in terms of its temporal and spatial evolution. Further, we provide a geometric relationship between the invariance parameters $\gamma$ and $\tau$. This relation is associated with the construction of the torus knot from the electric field maxima of the driving beam and a scaling factor. Our results offer an intuitive way to understand TKAM in high harmonic radiation and our explicit computations are in good agreement with previous findings \cite{Ballantine2016,Pisanty2019,Pisanty2019a}.
 
This paper is structured as follows. In Sec.~\ref{subsec:model} we discuss the geometric setup as well as Laguerre-Gaussian beams. Then, in Sec.~\ref{subsec:HHG}, we investigate the theoretical framework of HHG including the strong-field approximation and the quantum orbit approach. Moreover, in Sec.~\ref{subsec:BiCirc}, we discuss the symmetry transformations of twisted light and introduce the TKAM. Based on this framework, we present our results in Sec.~\ref{sec:discussion}. In Sec.~\ref{subsec:HHTemporalTorus}, we link the time delay parameter $\tau$ to the high harmonic radiation and determine $\tau$ explicit. This determination is based on the temporal and azimuthal evolution of the high harmonic radiation. Furthermore, in Sec.  \ref{subsec:HHSpatialTorus} we also link and determine the coordination parameter $\gamma$ to the high harmonic radiation. Considering the findings of Sec.~\ref{subsec:HHTemporalTorus} and \ref{subsec:HHSpatialTorus}, we finally demonstrate the geometric relation between $\tau$ and $\gamma$ in Sec.~\ref{subsec:2Dmap}. In the last section, we summarize the findings of this work.  

In the following, we will use atomic units ($\hbar=e=m_e=4\pi\epsilon_0=1$) unless stated otherwise.
\maketitle\section{\label{sec:methods}Theoretical Methods}
\maketitle\subsection{\label{subsec:model}Geometric setup and Laguerre-Gaussian beams}

\begin{figure*}[t!]
    \centering
    \includegraphics[width=\textwidth]{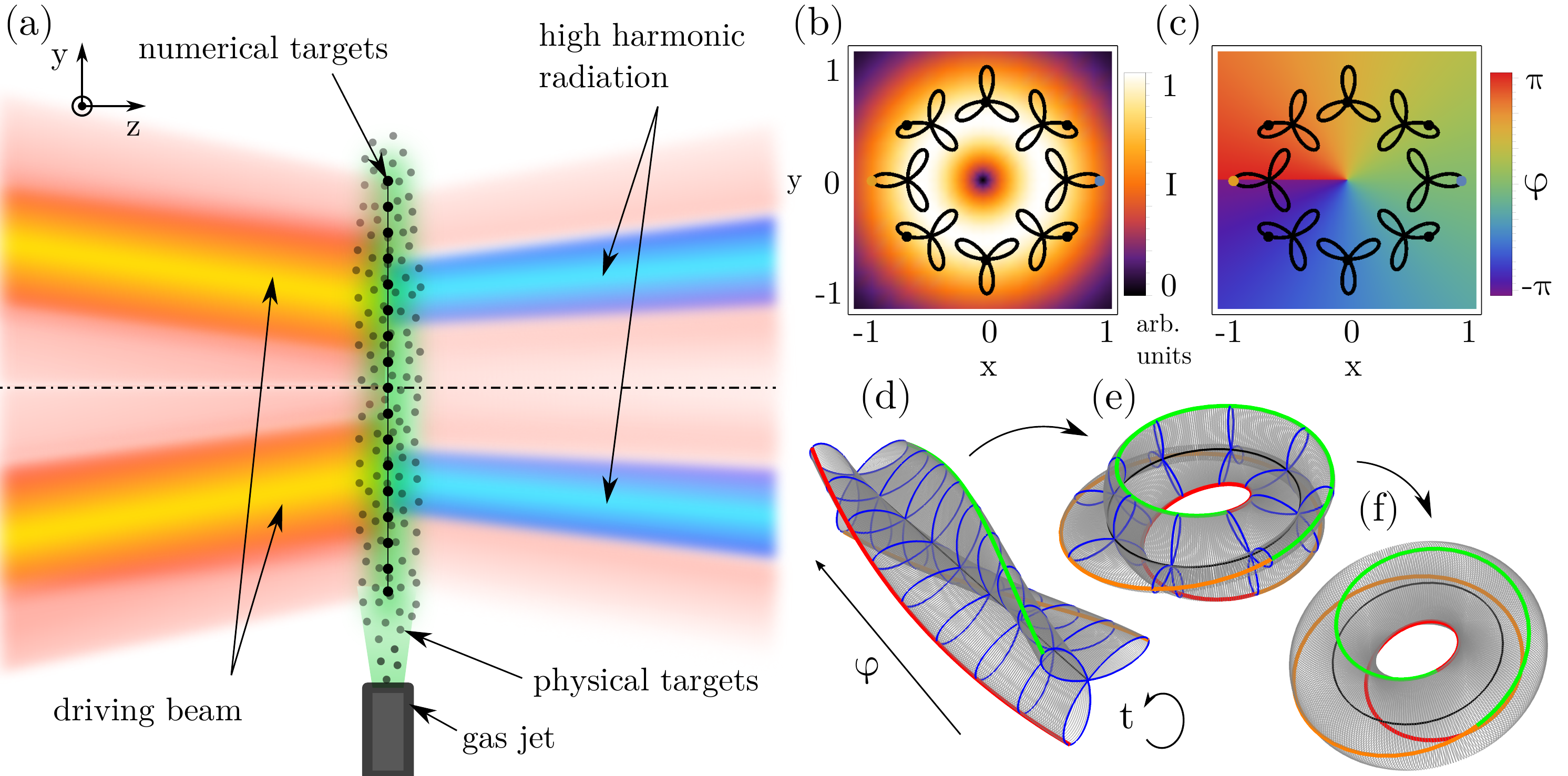}
    \caption{High-harmonic generation driven by bicircular LG beams. (a) Physical setup: A driving beam (orange), consisting of two superimposed LG beams, irradiates the physical target atoms (grey dots) from a gas jet (green) that is emitted perpendicularly to the optical axis (dash dot line). The target is approximated by a two-dimensional distribution of atoms in the $x-y-$plane (black dots). After the interaction of the target atoms with the driving beam, high-harmonic radiation (blue) is emitted from the interaction region. (b) Intensity distribution of the driving beam ($\ell_1=\ell_2=1,\;\omega_1=\omega_2/2=\omega$) in the target plane. Lissajous figures indicate the polarization of the superimposed LG beams for fixed azimuthal angles $\varphi$. The black dots indicate the position of the electric field vector for a fixed time $t$. The colored dots mark the electric field maxima for fixed $\varphi$ and $t$. (c) Phase of the electric field of the driving beam, together with the Lissajous figures also shown in (b). (d) The blue Lissajous figures are the same as in subfigure (b) and (c) where $\varphi=0$ denotes the position where the electric field vector is maximized (blue dot). The green, red and orange lines connect the positions of the field maxima. (e) If the open ends in (d) are connected, the green, red and orange lines that follow the field maxima are joined together and the resulting curve forms a torus knot.(f) Projection of the torus knot constructed in (e) onto a torus.}
    \label{setup}
\end{figure*}

We consider a two-dimensional planar atomic gas target, localized at $z=0$ and perpendicular to the optical axis $\textbf{e}_z$, see Fig.~\ref{setup}(a).
The gas target is irradiated by two superimposed counter-rotating bi-harmonic Laguerre-Gaussian (LG) beams. In paraxial approximation, their respective vector potentials $\bm{A}_{\ell_1,0}^{\omega_1\ \bm{\xi}_1}$ and  $\bm{A}_{\ell_2,0}^{\omega_2\ \bm{\xi}_2}$ in Coulomb gauge are given by \cite{Andrews2012}
\begin{align}
    \bm{A}_{\ell,n}^{\omega\,\bm{\xi}}(\textbf{r},t)=&-\frac{\sqrt{I_0}}{\omega}C_{\ell n}\frac{w_0}{w(z)}\left( \frac{r\sqrt{2}}{w(z)}\right)^{\mid\ell\mid}\mathrm{L}_n^{\mid\ell\mid}\left(\frac{2r^2}{w^2(z)}\right)\nonumber\\
    &\times \mathrm{exp}\left(-\frac{r^2}{w^2(z)}-i\left(k\frac{r^2}{2R(z)}+\ell\varphi-\psi (z)\right)\right)\nonumber\\&\times\mathrm{exp}\left(i(kz-\omega t)\right)\bm{\xi},\label{VP}
\end{align}
where $r,\varphi,z$ are cylindrical coordinates, $n$ is the radial index which determines the number of radial nodes and $\ell$ is the orbital angular momentum (OAM). Note, all mentioned angular momenta in this work are projected onto the optical axis unless stated otherwise. In the above expression, $\bm{\xi} \in \Big\{ \frac{1}{\sqrt{2}}\begin{pmatrix} 1\\ i \end{pmatrix},\frac{1}{\sqrt{2}}\begin{pmatrix} 1\\ -i \end{pmatrix}\Big\}$ is the polarization vector, in the following denoted with $\{\circlearrowleft,\circlearrowright\}$ respectively, $w_0$ is the waist radius, $k$ is the linear momentum and $\mathrm{L}_p^{\mid\ell\mid}$ are the generalized Laguerre polynomials. Here, $z_R=k\frac{w_0^2}{2}$ is the Rayleigh range, $I_0$ is the beam intensity at the maximum, $C_{\ell p}=\sqrt{\frac{2p!}{\pi (\pi+\mid\ell\mid)!}}$ is a normalization constant and $w(z)=w_0\sqrt{1+\frac{z^2}{z_R^2}} $ is the radius at which the field amplitude decreased to $1/e$, with the beam waist $w_0$. The radius of curvature is defined as $R(z)=z\left[ 1+\left(\frac{z_R}{z}\right)^2\right]$ and the Gouy phase as $\Psi (z)=\arctan{\left(\frac{z}{z_R}\right)}$. With the above vector potential, the electric field of the Laguerre-Gaussian beam can be computed with
\begin{align}
    \bm{E}_{\ell,p}^{\omega\,\bm{\xi}}(\bm{r},t)=-\partial_t\bm{A}_{\ell,p}^{\omega\,\bm{\xi}}(\bm{r},t).\label{ElectricFieldLG}
\end{align}
The frequencies, orbital angular momenta and polarizations of the beams $\bm{A}_{\ell_1,0}^{\omega_1\ \bm{\xi}_1}(\bm{r},t)$, $\bm{A}_{\ell_2,0}^{\omega_2\ \bm{\xi}_2}(\bm{r},t)$ are denoted as ($\omega_1$,$\omega_2$), ($l_1$,$l_2$) and ($\bm{\xi}_1,\bm{\xi}_2$), respectively. In order to simplify the notation, we will drop the indices $\omega,l,\bm{\xi}$ and write the vector potentials as well as the electric fields of the two beams simply as $\bm{A}_1(\bm{r},t),\bm{A}_2(\bm{r},t)$ and $\bm{E}_1(\bm{r},t),\bm{E}_2(\bm{r},t)$, respectively.

As a result of their OAM, LG beams exhibit helical phase fronts and a phase singularity on the optical axis. The helical phase fronts are induced by the angle-dependent factor $e^{-i\ell\varphi}$ \cite{Bazhenov1990,Harris1994}. On the other hand, the SAM $s$ is associated with the (local) polarization of these beams, $\xi\in\{\circlearrowleft,\circlearrowright\} \Leftrightarrow s\in\{+1,-1\}$, for left - and right circular polarized beams, respectively \cite{Bliokh2015}. 

 In this work, we will always assume counter-rotating beams with $\omega_2=2\omega_1$ and with equal intensities.
\maketitle\subsection{\label{subsec:HHG}Quantum orbit approach to high-harmonic generation}
In order to theoretically describe the HHG process driven by strong laser beams, we make use of the model by Lewenstein et al. \cite{Lewenstein1994}. In this model, the properties of the high-harmonic radiation can be derived from the time-dependent dipole moment, 
\begin{align}
    \bm{D}(\bm{r}_0,t_r)=&-i\int_{-\infty}^{t_r} dt_i\int d^3p\; \bm{d}^{\dagger}(\bm{p}+\bm{A}(\bm{r}_0,t_r))\bm{E}(\bm{r}_0,t_i)\nonumber\\&\times\bm{d}(\bm{p}+\bm{A}(\bm{r}_0,t_i))e^{-iS(\bm{r}_0,\bm{p},t_i,t_r)}.\label{fullSFA}
\end{align}
Here, $\bm{d}(\bm{p})=\braket{\bm{p}|\hat{\bm{r}}| g}$ denotes the dipole matrix element of the bound-free transition of the active electron with $\bra{\bm{p}}$ as electron plane-wave continuum state with momentum $\bm{p}$ and ground state $\ket{g}$. 
In the dipole moment (\ref{fullSFA}), the spatial dependence of the driving beam enters only parametrically. Therefore, each target atom experiences a spatially constant vector potential as well as a constant electric field. The phase $S(\bm{r},\bm{p},t_i,t_r)$ denotes the classical action of the electron in the continuum,
\begin{align}
    S(\bm{r}_0,\bm{p},t_i,t_r)=\int_{t_i}^{t_r}dt''\left( \frac{1}{2}\left[ \bm{p}+\bm{A}(\bm{r}_0,t'') \right]^2+I_p \right),
\end{align}
 where $I_p$ is the ionization potential of the atomic target. In order to evaluate the dipole moment (\ref{fullSFA}), we make use of the saddle-point approximation to solve the highly oscillating time and momentum integrals. With this approximation, the dipole moment reduces to \cite{Lewenstein1994,Le2016,Milo2002},
\begin{align}
    \bm{D}(\bm{r}_0,\omega_q)=&-i\sum_s\sqrt{\frac{(2\pi i)^2}{\mathrm{det}(\bm{\Theta} '')}}\left( \frac{-2\pi i}{t^(s)_r-t^{(s)}_i} \right)^{\frac{3}{2}}\nonumber\\&\times\bm{d}^{\dagger}(\bm{p}^{(s)}+\bm{A}(\bm{r}_0,t^{(s)}_r))\bm{E}(t^{(s)}_i)\nonumber\\&\times\bm{d}(\bm{p}^{(s)}+\bm{A}(\bm{r}_0,t^{(s)}_i))\nonumber\\&\times e^{-i\Theta(\bm{r}_0,\bm{p}^{(s)},t^{(s)}_r,t^{(s)}_i)}\label{QO} 
\end{align}
with 
\begin{align}
    \Theta(\bm{r}_0,\bm{p}^{(s)},t^{(s)}_r,t^{(s)}_i)=S(\bm{r},\bm{p}^{(s)},t^{(s)}_r,t^{(s)}_i)-\omega_qt^{(s)}_r. \label{DipolePhase}
\end{align} 
Here, $\Theta ''$ is the Hessian matrix of the dipole phase $\Theta$ with respect to  $t^{(s)}_r$ and $t^{(s)}_i$. The saddle points are denoted as $s=(p_s,t^{(s)}_r,t^{(s)}_i)$ and are solutions to the saddle point equations
\begin{subequations}
\label{SaddlePoint}
\begin{align}
\bm{p}^{(s)}=&-\frac{1}{t^{(s)}_r-t^{(s)}_i}\int^{t^{(s)}_r}_{t^{(s)}_i}d\tau \bm{A}(\tau),\\
I_p=&-\frac{1}{2}\left(\bm{p}^{(s)}+\bm{A}\left( t^{(s)}_r\right)\right),\\
q\omega-I_p=&\frac{1}{2}\left(\bm{p}^{(s)}+\bm{A}\left(t^{(s)}_i\right)\right).
\end{align}
\end{subequations}
If we insert the combined vector potential $\bm{A}(r_0,t) = \bm{A}_1(r_0,t) + \bm{A}_2(r_0,t)$ and electric field $\bm{E}(r_0,t) = \bm{E}_1(r_0,t) + \bm{E}_2(r_0,t)$ of the superimposed LG beams into Eq.~(\ref{QO}), we obtain the dipole moment for a specific harmonic order $q$ with energy $q\omega$. The electric field of the emitted radiation is then simply given by
\begin{align}
      \bm{E}_{H}(\bm{r}_0,t)&=\int d\omega_q  \bm{D}(\bm{r}_0,\omega_q)\omega_q^2e^{i\omega_qt}\nonumber\\
      &=\sum_q\bm{D}(\bm{r}_0,\omega_q)\omega_q^2e^{i\omega_qt}.\label{HHelectric}
\end{align}
In the present case, the energy spectrum of the high-harmonic radiation is discrete, since the driving beam components are monochromatic. In the following, we will refer to the high harmonic electric field (\ref{HHelectric}) as high harmonics (HHs).
The times $t^{(s)}_i$ and $t^{(s)}_r$ can be interpreted in a quasi-classical picture as \textit{ionization time}  and \textit{recombination time}, respectively \cite{Sansone2004,Keldysh1965}. The saddle-point equations (\ref{SaddlePoint}) and the respective saddle points for these times imply a strong relationship between the magnitude of the driving beam's electric field and the dipole moment (\ref{QO}) \cite{Milo2000a,Milo2000b}.   
Therefore, this relation is imprinted in the temporal and azimuthal evolution of the field of the outgoing HHs. The evolution of a field $\mathrm{F}(x,y)$ with respect to the arbitrary variable $x$ describes the function value $\mathrm{F}(x,y_0)$ for increasing or decreasing values of $x$ within the domain of $x$. Here, $y_0$ denotes a set of fixed variables in a way according to which 
\begin{align}
    \mathrm{F}(x)\equiv\mathrm{F}(x,y_0).\label{DefEvolution}
\end{align}
The temporal and azimuthal evolution of the driving beam and the HHs can be likewise denoted with regard to their field maxima $\bm{E}_{\varphi}(t)=\mathrm{max}_{\varphi}\left(\bm{E}(r_0,\varphi,z_0,t)\right)$ and $\bm{E}_{t}(\varphi)=\mathrm{max}_t\left(\bm{E}(r_0,\varphi,z_0,t)\right)$. Here, the index $i=t,\varphi$ denotes the dimension $i$. The respective variables $r_0$ and $z_0$ are associated with the fixed variable set $y_0$ in Eq.~(\ref{DefEvolution}). This restriction from the general field $\bm{E}(\bm{r},t)$ to the corresponding field maxima is a simplification to display the crucial dynamical behavior concerning $t$ and $\varphi$.

The counter-rotating $\xi_1=\;\circlearrowleft$, $\xi_2=\;\circlearrowright$ composition of the bichromatic driving beam in Sec.~\ref{subsec:model} is necessary, since HHG with bicircular co-rotating beams is forbidden by SAM selection rules  \cite{Paufler2018,Pisanty2014}. In general, the LG beams $\bm{A}_1(\bm{r},t),\bm{A}_2(\bm{r},t)$  can possess arbitrary integer OAM and frequencies $\omega_i=p_i\omega$ with co-prime integers $p_1$ and $p_2$. Moreover, both LG beams share a common fundamental frequency $\omega$. With the above assumptions for the driving beam parameters, their combined electric field takes the form
\begin{align}
    \bm{E}(\varphi,t)\equiv&\bm{E}(r_0,\varphi,z_0,t)\nonumber\\ =& E_0(r_0,z_0)\nonumber\\&\times \begin{pmatrix} \cos(l_1\varphi-\omega_1t)+\cos(l_2\varphi-\omega_2t)\\
    \sin(l_2\varphi-\omega_2t)-\sin(l_1\varphi-\omega_1t)\\
    \end{pmatrix}.\label{electricfield}
\end{align}
The radial position $r_0 = \mathrm{const}.$ is chosen such that the local intensity of the driving beam is maximized. Thus, the coordinate space of the driving beam is reduced to two dimensions for a fixed position on the optical axis $z_0=\mathrm{const}.=0$ and $r_0$. The remaining electric field is a function which depends on the time $t$ and the azimuthal angle $\varphi$ ($\bm{E}(r_0,\varphi,z_0,t)\rightarrow\bm{E}(\varphi,t)$)

This reduced coordinate space of the driving beam (\ref{electricfield}) is found in the HHs (\ref{HHelectric}) as well. Here, the emitting targets form a ring with radius $r_0$ that is associated with the leading contribution of the high harmonic time-averaged intensity profile in the $x-y-$plane, see Fig.~\ref{setup}(b). The temporal evolution of the HHs exhibits $p_1+p_2$ separate sections in one period $T=2\pi/\omega$ of the fundamental frequency $\omega$. The sum $p_1+p_2$ represents the number of intensity maxima of the electric field, concerning the time, where each section is associated with one maximum. Here, a section is defined as the time interval between two intensity minima. An illustration can be found in Fig.~\ref{LSgraphic}(b). These separate sections are likewise known as attosecond pulse for one section and attosecond pulse trains (APTs) for multiple sections. Attosecend pulses are separated by points in time with vanishing electric field intensity. This indicates that the APTs mimic the temporal evolution of the electric field at the position of the target atom.

In the HHG process, the active electron accumulates the dipole phase (\ref{DipolePhase}). This dipole phase induces a constant phase shift between the electric field of the driving beam and the emission of the HHs. The induced phase shift is equal for all target atoms, since the ionization and recombination times are characteristic. Therefore, the HHs are emitted after a constant time delay with respect to the electric field of the driving beam. The phase shift, in general, does not affect our analysis below. Formally, this can be seen in Eq.~(\ref{electricfield}) where the azimuthal angle as well as the time appear as an argument of the trigonometric functions but not as an argument of the electric field amplitude.

\maketitle\subsection{\label{subsec:BiCirc}Bicircular Laguerre-Gaussian beams and TKAM}
As discussed above, monochromatic LG beams are eigenstates of the OAM and SAM projection operators. Therefore, these quantities are conserved during the propagation of the beams. According to Noethers theorem, the conservation of a quantity is  the system's invariance under a global continuous symmetry transformation \cite{Noether2011}. The conservation of OAM and SAM is the consequence of transformations which leave the electric field of a LG beam invariant. The transformations associated with the conservation of OAM are rotations of the coordinate system with respect to the optical axis, i.e. the angle $\varphi$. Likewise, SAM conservation is associated with the rotation of the electric field vector itself. In addition, each of these rotations need to be associated with a time translation to fulfill the symmetry conditions. Furthermore, the subsequent application of both rotations is associated with the conservation of total angular momentum (TAM). These rotations are called coordinated rotations (CRs). The corresponding symmetry transformations related to the respective conservation laws are explicitly denoted as,
\begin{subequations}
\label{invMono}
\begin{align}
\bm{E}_i(\bm{R}^{-1}(\varphi)\bm{r},t) &= \bm{E}_i(\bm{r},t+\frac{\ell_i\varphi}{\omega_i}),\label{invOAM}\\
\bm{R}(\gamma\varphi)\bm{E}_i(\bm{r},t) &= \bm{E}_i(\bm{r},t\pm\frac{\gamma_i\varphi}{\omega_i}),\label{invSAM}\\
\bm{R}(\gamma\varphi)\bm{E}_i(\bm{R}^{-1}(\varphi)\bm{r},t) &= \bm{E}_i (\bm{r},t+(\ell_i\pm\gamma_i)\frac{\varphi}{\omega_i}).\label{invTAM}
\end{align}
\end{subequations}
In these expressions, $\bm{R}(\gamma_i\varphi)$ and $\bm{R}^{-1}(\varphi)$ represent a rotation of the electric field and the coordinate system, respectively. The variable $\gamma_i\varphi$ is associated with the orientation of the Lissajous figure Fig.~\ref{setup}(b) in the transverse polarization plane. The parameter $(\pm)\gamma_i$ represents the coupling parameter for CRs and the sign $\pm$ denotes the polarization. For the case of monochromatic LG beams, $\gamma_i$ can take an arbitrary value in the symmetry transformations (\ref{invMono}).

Equations (\ref{invMono}) state that OAM, SAM, and TAM are well-defined quantum numbers and are therefore independently measurable. Equivalently, monochromatic LG beams can be classified by their respective symmetry group. This group is called the full symmetry group of paraxial optics $\textrm{SO}(2)\times \textrm{SO}(2)$ \cite{Pisanty2019a,Bandres2009}.

We now turn to the more intricate case of bicircular twisted light beams. In general, the superposition of two LG beams does not need to fulfill the symmetry properties of individual components. This property can be seen explicitly if we focus on the temporal and azimuthal dependencies of monochromatic LG beams in Eqs. (\ref{invMono}). Here, the symmetry transformations are crucially dependent on the OAM $\ell_i$, the frequency $\omega_i$, and the polarization $\bm{\xi}_i$. The superposition of an additional beam with different parameters $\ell_1\neq\ell_2,\;\omega_1\neq\omega_2$ and/or $\bm{\xi}_1\neq\bm{\xi}_2$ will lead to a violation of the symmetry conditions of the first beam. This symmetry breaking induces ill-defined OAM, energy, and/or SAM eigenvalues of the combined beam.
Moreover, for different beams, the TAM is also not well-defined, $\hat{J_z}\ket{E_i}\neq j_z\ket{E_i} $, which restricts the beam symmetries even further.

In the vector potentials (\ref{VP}), we see that the temporal and azimuthal contributions to the complex phase are identical besides a constant factor. This can be recognized even more easily in the arguments of the trigonometric functions in the combined electric field (\ref{electricfield}). Here, the time $t$ and the azimuthal angle $\varphi$ are coupled to the frequency $\omega_i$ and the OAM $\ell_i$, respectively. These parameters uniquely define the symmetric properties of a single monochromatic LG beam. The frequency, as well as the OAM of the monochromatic LG beam, define the temporal and azimuthal evolution with $\omega_it$ and $\ell_i\varphi$, respectively. Thus, the temporal evolution of two superimposed LG beams is not determined by either $\omega_1$ or $\omega_2$ alone. This holds for the azimuthal evolution and $\ell_1$ or $\ell_2$ as well. The reason for this behavior lays in the nonlinearity of the trigonometric functions of the electric field (\ref{electricfield}). Here, the different arguments of the trigonometric functions do not allow to rewrite the superimposed beam in a form similar to a single LG beam. Thus, it is not possible to define a proper symmetry transformation under which the superimposed beam is invariant. Moreover, the explicit parameters $\ell_i$ and $\omega_i$ of the superimposed beam define the symmetry transformations in Eq.~(\ref{invMono}). Therefore, the explicit symmetry transformations need to be adjusted to a particular beam.

The electric field of monochromatic bicircular twisted light beams can be expressed as
\begin{align}
    \bm{E}=E_1e^{i\ell_1\varphi}\bm{\xi}_1+E_2e^{i\ell_2\varphi}\bm{\xi}_2.\label{monochromaticElectric}
\end{align}
Here, we omit the spatial and temporal dependencies of the amplitudes $E_i=E_i(r,z,t)$. Ballantine et al. \cite{Ballantine2016} found that such a field satisfies the eigenvalue equation 
\begin{align}
    (\hat{L}+\gamma \hat{S})\bm{E}=j_{\gamma}\bm{E},
\end{align}
with the eigenvalue
\begin{align}
    j_{\gamma}=\frac{\ell_2+\ell_1}{2}
\end{align}
 and the parameter
\begin{align}
    \gamma=\frac{\ell_2-\ell_1}{2}.
\end{align}
Thus, a half-integer TAM quantization arises for monochromatic bicircular beams \cite{Ballantine2016}.

However, it is possible to generalize the idea of Ballantine et al. to bichromatic bicircular beams. In this case, the previously arbitrary parameters $\gamma_i$ need to be restricted. The restriction intends to induce the same rotation of the electric field for both beams. This can be written in an explicit way as
\begin{align}
    \gamma_1\varphi&\equiv\gamma_2\varphi\nonumber\\\Rightarrow\;\gamma_1&\equiv\gamma_2,\nonumber
\end{align}
where $\gamma_i\varphi$ is the angle of rotation of the electric field. The restriction enables a description of the driving beam which fulfills a symmetry transformation similar to Eq.~(\ref{invTAM}).

The solution to these symmetry transformation issues was found by Pisanty et al. \cite{Pisanty2019a}. His solution is a generalization of the findings from Ballentine et al. \cite{Ballantine2016} and correspond to a new form of angular momentum called TKAM \cite{Pisanty2019a,Pisanty2019}. The symmetry transformation for bicircular bichromatic twisted light beams are then
\begin{align}
    \bm{R}(\gamma\varphi)\bm{E}(\bm{R}^{-1}(\varphi)\bm{r},t)=\bm{E}(\bm{r},t+\tau\varphi),\label{CRsSym}
\end{align}
where $\tau$ is the time delay parameter associated with the coordination parameter $\gamma$ on the left-hand side. Most importantly, bicircular bichromatic twisted light beams do not fulfill any symmetry conditions concerning the rotation of the coordinate system or the electric field separately. Thus, OAM and SAM are not well defined for those superimposed beams.

The TKAM operator is then defined as
\begin{align}
    \hat{J_{\gamma}}=\hat{L}+\gamma\hat{S}.\label{opTKAM}
\end{align}
Contrary to the arbitrary $\gamma_i$ of Eq.~(\ref{invTAM}), the superimposed driving beam obeys Eq.~(\ref{CRsSym}) only for discrete values of $\gamma$,
\begin{align}
    \gamma=\frac{\ell_2\omega_1-\ell_1\omega_2}{\omega_1+\omega_2}=\frac{\ell_2p_1-\ell_1p_2}{p_1+p_2}\label{gamma}.
\end{align}
Here, $\gamma$ is defined to be equal for both LG beams and fulfills Eq.~(\ref{CRsSym}). These restrictions reduce all possible coordination parameters $\gamma_i$ of the LG beams to a set of integer multiples of $\gamma$. Therefore, $\gamma$  defines the symmetry transformations of the superimposed driving beam. The time delay parameter $\tau$ on the RHS of Eq.~(\ref{CRsSym}) classifies the time shift induced by a CR about $\varphi$ and $\varphi\gamma$.

Thus, the symmetry group of the driving beam is a subgroup of the full symmetry group of paraxial optics $\mathrm{SO}(2)\times\mathrm{SO}(2)$.     
Since the full symmetry group can (geometrically) be interpreted as a tensor product of two circles, it constructs a flat torus Fig.~\ref{setup}(f) (grey surface).
While, in general, tori of higher genus exist and are mathematically well known, only a flat torus of genus $g=1$ (one hole) obeys the geometric properties of the symmetry group $\mathrm{SO}(2)\times\mathrm{SO}(2)$. The surface of these tori can be parametrized by two (locally) orthogonal dimensions, the poloidal dimension and the toroidal dimension. The toroidal dimension describes the rotation around the hole, while the poloidal dimension locally rotates orthogonally.

In general, two additional dimensions exist in our case. However, we can reduce these by fixing the poloidal- and toroidal radius $R^{pol}\equiv R^{pol}_0,R^{tor}\equiv R^{tor}_0$ with $R^{pol}_0,R^{tor}_0\neq 0$. The remaining torus surface is two-dimensional. This assumption is physically imprinted in the restriction to the target ring of radius $r_0$ in Eq.~\ref{electricfield} ($R^{pol}_0=r_0$). Furthermore, it is imprinted in the finite time interval from $t\in\left[t_0,t_0+T\right]$ ($r^{tor}_0=T/2\pi$).

Finally, since Eq.~(\ref{CRsSym}) represents a subgroup of the full group of paraxial optics, this subgroup is located within the torus as well. If we consider discrete $\tau,\gamma$ and constant $r^{pol}_0,r^{tor}_0$ the full four-dimensional torus of paraxial optics is further reduced to only one dimension. Therefore, the subgroup associated with the dynamic symmetry of the driving beam (\ref{CRsSym}) is one-dimensional. Each element of this one-dimensional subgroup winds around the surface of the two-dimensional torus. The elements of the subgroup are called torus knots and are characterized by two integer numbers $m,n$ \cite{Adams2004}. The integers $m,n$ count the number of times the knot crosses a fixed point with respect to the poloidal or toroidal direction, respectively. Note that $m$ is a signed integer such that these numbers build a $(m,n)$ signed torus knot. The sign of $m$ denotes the orientation of the rotation about the poloidal axis.

The torus knot of the driving beam is illustrated in Fig.~\ref{setup}(f) with $\omega_1=\omega$, $\omega_2=2\omega$ and $\ell_1=\ell_2=1$.
For every fixed azimuthal angle $\varphi_0$ (toroidal direction), there are $p_1+p_2$ electric field maxima with respect to time. These maxima are associated with the trefoil (poloidal direction) tips of the Lissajous figure in Fig \ref{setup}(d). If the azimuthal angle is varied, the trefoil tips describe lines. These lines are the electric field maxima depending on time $t$ and the azimuthal angle $\varphi$. The poloidal dimension is associated with the time $t\in \left[ t_0,t_0+T\right]$, where $T = 2\pi/\omega$ is the period associated to the fundamental frequency. Counting the poloidal and toroidal rotations of the torus knot in Fig.~\ref{setup}(f) leads to the torus knot $(m=-1,n=3)$ which corresponds to the coordination parameter $\gamma=-\frac{1}{3}$.

The associated time delay parameter $\tau$ in Eq.~(\ref{CRsSym}) then reads
\begin{align}
    \tau=\frac{\ell_1+\ell_2}{p_1+p_2}.\label{tau}
\end{align}
A comparison of Eq.~(\ref{gamma}) and Eq.~(\ref{tau}) reveals the nature of their dependencies on the driving beam parameters $\omega_i,\ell_i$. The beam parameter dependency agrees with the discussion at the beginning of this subsection.
Moreover, the TKAM eigenvalue can be rewritten as a function of the time delay parameter $\tau$,
\begin{align}
     j_{\gamma}^{(q)}=q\omega\tau.\label{gammatau}
\end{align}
In the past, different methods have been developed to classify HHs from bicircular LG beams \cite{Paufler2018,Paufler2019}. These methods predominantly classify the HHs within photon-counting methods. The counting methods distinguish between photons of different beams implicitly. Therefore, the driving beam, as well as the HHs, are treated as two separate beams. This treatment prohibits the construction of a well-defined angular momentum operator.\\
The TKAM, on the other hand, approaches the bicircular driving beam as a single beam and leads to mathematically well-defined eigenvalues with the full driving beam as the corresponding eigenstate. Therefore, the eigenvalue of the TKAM operator acts as a \textit{good} quantum number that is conserved in the HHG process and increases linearly with the harmonic order $q$, see Eq.~(\ref{gammatau}). 

Before we continue with the discussion of explicit results, let us briefly highlight the most important points of this section. The temporal and azimuthal evolution of the HHs and the driving beam are imprinted in their time and azimuthal angle-dependent field maxima. Therefore, the HHs and the driving beam can be reduced to one-dimensional lines, or knots, which represent the crucial properties of these fields.

In contrast to a monochromatic driving beam, bichromatic bicircular driving beams cannot be characterized by their projected TAM $j_z$. Moreover, these driving beams do not fulfill the dynamical symmetry transformation Eq.~(\ref{invTAM}). Instead, the explicit transformations are coupled to the driving beam parameters ($\omega_1,\omega_2$) and  ($\ell_1,\ell_2$) itself. Therefore, the superposition of different beams leads naturally to a breaking of the dynamical symmetries associated with OAM, SAM, and TAM. This classification issue is solved with the introduction of the TKAM operator. The TKAM operator leads to a reduced version of the dynamical symmetry concerning the LG beams. This reduced dynamical symmetry is realized within the discrete parameters $\gamma$ and $\tau$ of the former continuous parameters of the LG beams $\gamma_i$ and $\tau_i$.

\maketitle\section{\label{sec:discussion}Discussion and Results}
In this section, we will demonstrate how the time delay parameter $\tau$ and the coordination parameter $\gamma$ can be associated with a torus knot (Secs.~\ref{subsec:HHTemporalTorus} and \ref{subsec:HHSpatialTorus}) and derive a method to transform the $\tau$ torus knot into a $\gamma$ torus knot (Sec.~\ref{subsec:2Dmap}). Within this method, we associate both tori with the same two-dimensional representation of either the electric field maxima of the driving beam or the HHs.    
\maketitle\subsection{\label{subsec:HHTemporalTorus}Classification of HHs based on their temporal evolution}
\begin{figure*}[t!]
    \centering
    \includegraphics[width=\textwidth]{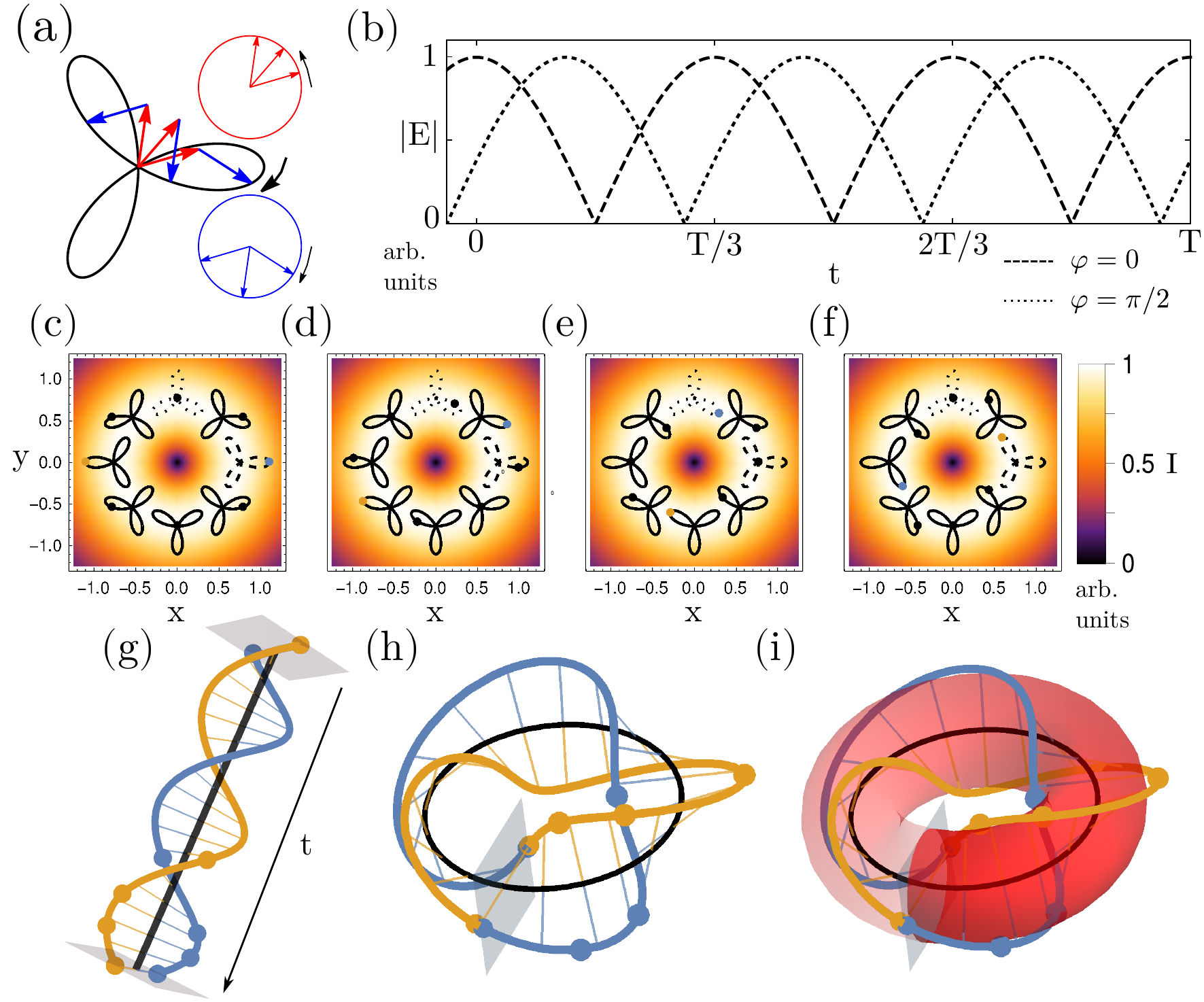}
    \caption{
    (a) Bicircular field as a superposition of a circularly polarized (red) and its counter-rotating second harmonic (blue), which gives rise to the Lissajous figure (trefoil shape) if the position of the electric field vector is followed in time.
    (b) Absolute values for the electric field at azimuthal angles $\varphi = 0$ and $\varphi=\pi/2$ as a function of time.
    (c)-(f) Orientation of the Lissajous figures for various values of the azimuthal angle $\varphi$ of a bicircular driving beam ($\ell_1=\ell_2=1,\;\omega_1=\omega_2/2=\omega$). The background shows the cycle averaged intensity distribution of the driving beam with the characteristic singularity on the beam axis. Each subfigure (c)-(f) shows the electric field at the target for increasing times $t=0,\,T/12,\,T/6,\,T/3$. The black dots indicate the electric field at the respective angle in the target plane. The colored dots denote the maxima of the electric field concerning the azimuthal angle similar to Fig.~\ref{setup}(b) and (c).
   (g) Temporal evolution of the HHs with respect to the black axis (time axis). The separate spirals indicate the HH intensity maximum with respect to the azimuthal angle at the target (light gray plane at the bottom). The dots are the field maxima that we also show in the top row.
    (h) The time axis in (g) is bent and connected for the times $t=0$ and $t=T$.
    (i) The resulting closed line of intensity maxima has the topology of a torus knot. }
    \label{LSgraphic}
\end{figure*}
\begin{figure*}[t!]
    \centering
    \includegraphics[width=\textwidth]{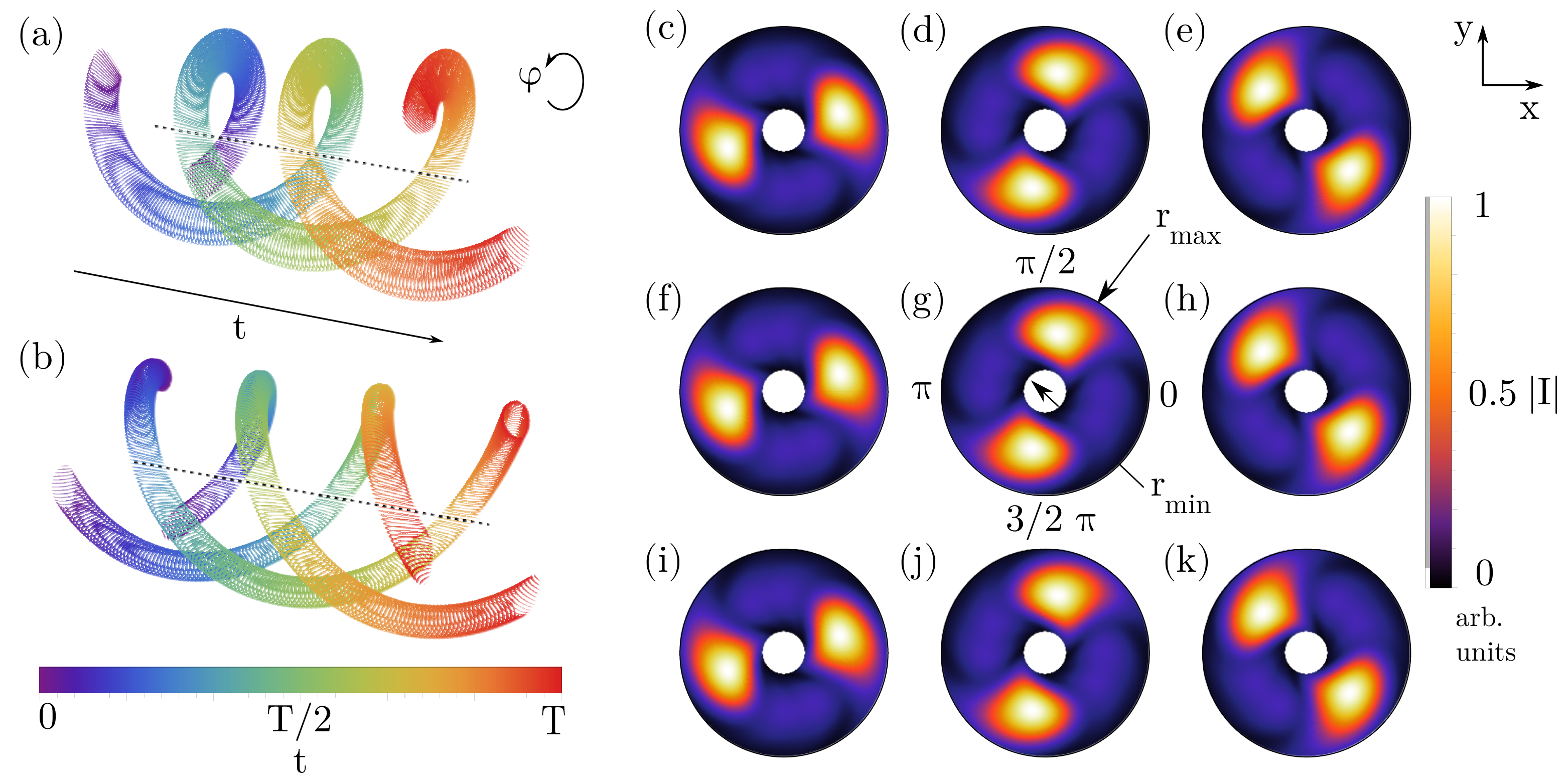}
    \caption{(a),(b) Temporal evolution of the light springs formed by the HHs. The figures display the contour obtained at 80 per cent of the maximum electric field strength. In order to compute the time signal, we included the $16th$ up to the $25th$ harmonic. The parameters of the driving beam are (a) $\ell_1=\ell_2=1,\;\omega_1=\omega_2/2=\omega$ and (b) $\ell_1=2,\;\ell_2=1,\;\omega_1=\omega_2/2=\omega$. (c)-(k) Intensity distribution of (a) for the times $t=j\frac{T}{9}$ with the integers $j\in[1,9]$, so that $t$ increases from (c) $t=\frac{T}{9}$ to (k) $t=T$ }
    \label{temporalTorusNumeric}
\end{figure*}
In our explicit calculations below, we will fix the beam intensities ${I_0}_1={I_0}_2=4\times10^{14}\;\mathrm{W/m^2}$, the fundamental frequency $\omega=0.057$, the beam waist at the focus, $w_0=60\;\mathrm{\mu m}$, the atomic ionization potential $I_p=15.75\;\mathrm{eV}\;(\mathrm{Ar})$, the explicit frequencies of the LG beams $\;p_1=1,\;p_2=2$, and the OAM of the LG beams $l_1=1,\;l_2=1$ unless stated otherwise.

The coordination parameter $\gamma$ of the driving beam is associated with a torus knot, as shown in Fig.~\ref{setup}(d)-(f) \cite{Pisanty2019a}. Therefore, the azimuthal dimension corresponds to the toroidal axes of the torus and the time on the poloidal axes. The physical dimension which is associated with the toroidal axis determines the torus knot as either a \textit{spatial torus knot} or a \textit{temporal torus knot}. Accordingly, Fig.~\ref{setup}(f) shows a spatial torus knot since the azimuthal angle is associated with the toroidal direction. 

On the other hand, the time delay parameter $\tau$ in the symmetry transformation (\ref{CRsSym}) of the electric field can also be related to a torus knot.
In Sec.~\ref{subsec:model}, we demonstrated that the intensity maxima of the HHs and the electric field of the driving beam have a close relationship. This relationship is physically imprinted as a time delay $\Delta t=t^{(s)}_r-t^{(s)}_i$ between the ionization time of a specific target atom and the emission of a respective harmonic photon. 
Based on the symmetry transformation (\ref{CRsSym}), the time delay can also be interpreted as a CR. The explicit angle of rotation $\varphi$ then reads
\begin{align}
    \Delta t =\tau\varphi_{\Delta t}\;\;\Rightarrow\;\; \varphi_{\Delta t} \equiv\frac{\Delta t}{\tau}.\label{DelayHHG}
\end{align}
In the numerical calculations below, we restrict the sum over all saddle points in the dipole moment (\ref{QO}) to the saddle point associated with the leading contribution. In general, this saddle point is related to the shortest trajectory in the quantum orbit approach \cite{Milo2019}. The finite travel time of the continuum electron leads to a finite time delay $\Delta t$ between ionization of the active electron and emission of a harmonic photon. Each saddle point has characteristic ionization and recombination times which induce the constant time delay. In the first instance, the constant time delay can be recognized as the angle $\varphi_{\Delta t}$ between the intensity maximum of the driving beam and the intensity maximum of the HHs for a fixed time. Note that the respective intensity maxima of the driving beam and the HHs evolve identically with $t$ and $\varphi$ up to the constant azimuthal shift.

Let us now focus on the temporal and azimuthal evolution of the HHs. As other authors have already shown \cite{Rego2019}, HHs generated from bicircular driving beams exhibit a spiral intensity distribution \cite{Garcia2013}. Due to their geometry, these spiral structures are called \textit{light springs}. 

In the following, the origin and shape of these light springs will be discussed in more detail. In Fig.~\ref{LSgraphic}(c)-(f), we present the intensity distribution of the driving beam ($\ell_1=\ell_2=1;\;\omega_1=\omega_2/2=\omega$) near the focus. The small Lissajous figures demonstrate how the orientation of the local electric field changes along the azimuthal angle. Each Lissajous figure denotes the electric field vector as it runs over one period $T$ of the fundamental frequency.
The black dots indicate the electric field at $t=0$. Figure \ref{LSgraphic}(b) shows the modulus of the electric field as a function of time for different values of the azimuthal angles. For $\varphi=0$, the electric field is maximized at $t=0$, which is also indicated by the blue dot in (c). At different azimuthal angles, the electric field has different values for $t=0$, as demonstrated by the position of the black dot in the Lissajous figures which changes along the azimuthal angle.
According to Eq.~(\ref{CRsSym}), a change of $\varphi_0$ in the azimuthal direction causes a delay in time of $\tau\varphi_0$.
As discussed in Sec.~\ref{subsec:model}, the leading contribution of the HHs is emitted from a narrow region around the modulus of the electric field maximum.

In general, HHG driven by bicircular counter-rotating beams leads to the emission of $p_1+p_2$ separate attosecond pules per fundamental period $T$. Thus, the maxima of the APT are separated in time by $\frac{2\pi}{(p_1+p_2)\omega}$.
Furthermore, a change of the azimuthal angle of the superimposed beam leads to a time delay of $2\pi\frac{\ell_1+\ell_2}{(p_1+p_2)\omega}$. Consequently, there are exactly $\ell_1+\ell_2$ maxima along the azimuthal direction. These $\ell_1+\ell_2$ maxima simultaneously lead to the emission of an attosecond pulse which ensures that the light spring consists of $\ell_1+\ell_2$ spirals.

This behavior is visualized in Fig.~\ref{LSgraphic}(g) for the specific parameters $\ell_1=\ell_2=1,\;\omega_1=\omega_2/2=\omega$. If the open ends of the light spring are connected similarly as shown in Fig.~\ref{setup}(c), they form the torus knot displayed in Fig.~\ref{LSgraphic}(h). As before, this knot can be projected onto the surface of a torus, which is shown in Fig.~\ref{LSgraphic}(i). Again, this torus knot can be classified by the winding number around the toroidal and the poloidal axes, respectively. Thus, the torus knot visualized in Fig.~\ref{LSgraphic}(i) is a ($m=3,n=2$) torus knot. We designate this knot as a \textit{temporal torus knot}, in analogy to its spatial counterpart above.\\

If we compare the numbers $m=3$ and $n=2$ to the general scenario outlined in the previous two paragraphs, we can deduce the dependencies

\begin{subequations}
\begin{align}
    n&=\ell_1+\ell_2,\label{TemporalTorusN}\\
    m&=p_1+p_2.\label{TemporalTorusM}
\end{align}
\end{subequations}
These relations allow to determine the time delay parameter on the basis of the geometric representation of the light spring,
\begin{align}
    \tau=\frac{n}{m\omega}=\frac{\ell_1+\ell_2}{\omega_1+\omega_2}.
\end{align}

For the special torus knot depicted in Fig.~\ref{LSgraphic}(g)-(i), the time delay parameter thus follows as $\tau=\frac{2}{3}T$.

Note that Eqs.~(\ref{TemporalTorusN}) and (\ref{TemporalTorusM}) contain no information about the respective parameters $l_i$ and $p_i$ of the two individual LG beams. However, this is not surprising since the TKAM treats the driving beam as a single object. Therefore, the driving beam itself is not characterized by single parameters of the driving beam more than it is by the parameters that characterize their composition.

With this geometric relation between $\tau$ and the HHs, the time delay parameter can be simply read off of the light spring. Subsequently, the eigenvalue of the TKAM operator can be calculated without knowledge of the explicit initial beam configuration via
\begin{align}
    j_{\gamma}^{(q)}=q\frac{n}{m}.
\end{align}
We present explicit numerical results in Fig.~\ref{temporalTorusNumeric}. Here, the driving beam parameters are (a) $\ell_1=\ell_2=1,\;\omega_1=\omega_2/2=\omega$ and (b) $\ell_1=2,\;\ell_2=1,\;\omega_1=\omega_2/2=\omega$ and the harmonic order ranges from $16$ to $25$. These plots are in excellent agreement with the properties of the light springs introduced above. Figures \ref{temporalTorusNumeric}(a) and (b) display two and three spirals, respectively, which agree with our discussion above. Figure~\ref{temporalTorusNumeric} (c)-(k) illustrates the intensity distribution of the HHs for uniform time steps. Note that, for every point in time, the number of spirals is $\ell_1+\ell_2=2$ in (a) and (c)-(k),  and $\ell_1+\ell_2=3$ in (b). The intensity distributions are identical for three separate times, which is linked to the fixed number $p_1+p_2=3$ of temporal maxima. Thus, this representation contains no information about the polarization of the HHs.

\maketitle\subsection{\label{subsec:HHSpatialTorus}Classification of HHs based on their spatial evolution}
\begin{figure*}[t!]
    \centering
    \includegraphics[width=\textwidth]{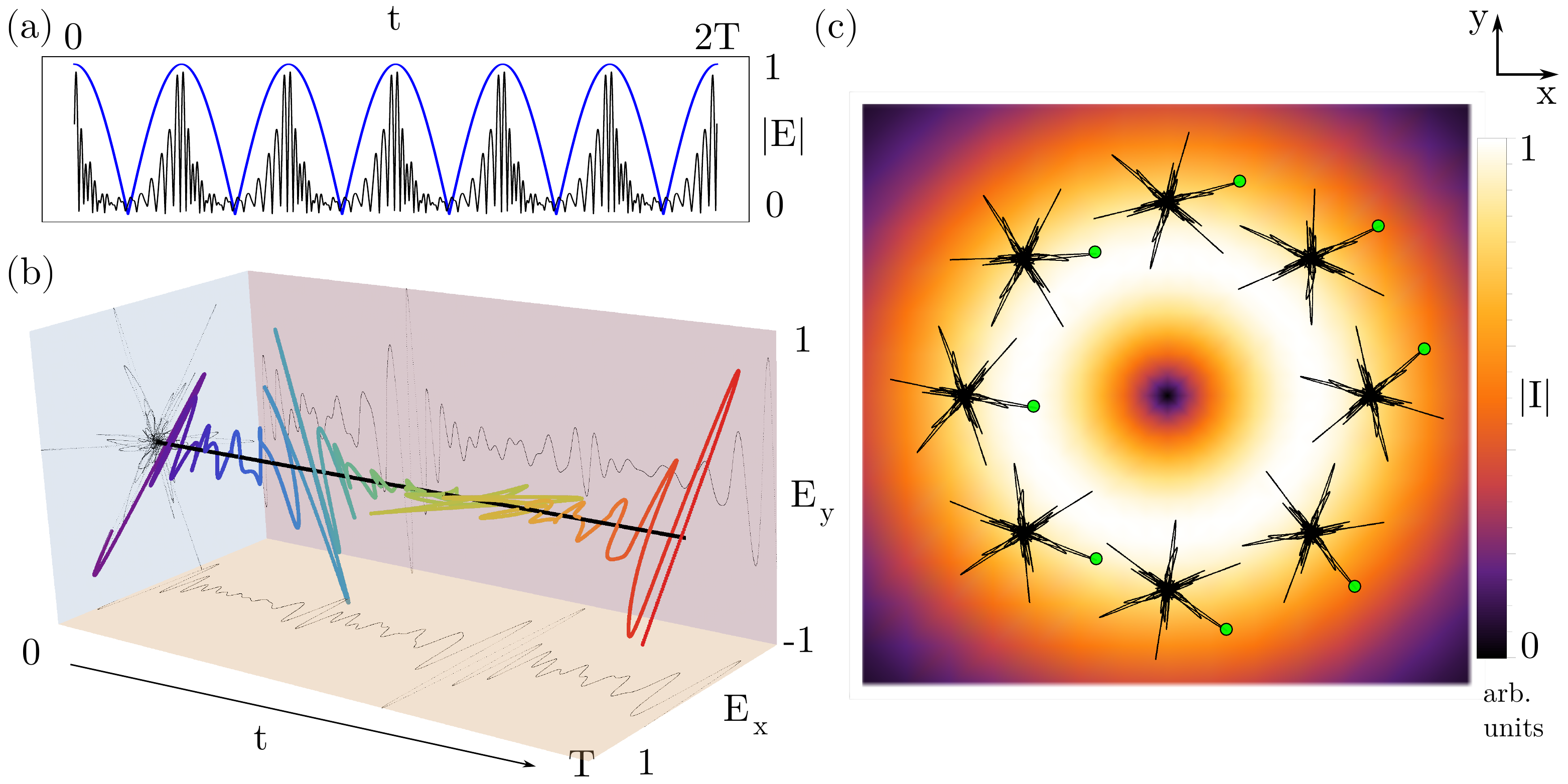}
    \caption{ (a) Absolute value of the electric field of the driving beam (blue) and the superimposed electric fields of the HHs at a fixed position (black). The changing color indicates the temporal evolution. (b) Temporal evolution of the HHs and their associated two-dimensional projections. (c) Spatial evolution of the emitted HHs with resolved angle-dependent polarization. The polarization figure is also shown in (b) as projection onto the $E_x - E_y -$plane.}
    \label{azimuthalTorus}
\end{figure*}

In the previous subsection, the light spring was associated with the time delay parameter $\tau$. Here, we show that the coordination parameter $\gamma$ can similarly be associated with the HHs. The method to do so is similar to the one illustrated in Fig.~\ref{setup}(d)-(f). In contrast to Fig.~\ref{setup}(d)-(f), however, there is no analytic formula that describes the time-dependent HH maxima. Fortunately, the explicit times $\bar{t}$ that maximize the HHs in the saddle-point approximation (\ref{QO}) of the dipole moment need not be known explicitly. As discussed in Sec.~\ref{subsec:HHG}, the time delay $\Delta t$ between the ionization and recombination times of the active electron remain constant if the recombination time and azimuthal angle are varied, respectively. Again, this constant time delay allows to couple each electric field vector of the driving beam to an unknown HH dipole vector. Also, identical electric field strengths induce identical HH dipole strengths which induces only a difference in the polarization. Thus, even if the explicit dependency is not known, each HH dipole maximum can be associated with an electric field maximum.

Moreover, since HHG is a parametric process, the HHs need to preserve the symmetric structure of the electric field with respect to time \cite{Milo2000a}. Therefore, since the electric field of the driving beam has $p_1+p_2$ field maxima within one period of the fundamental frequency $T$, the polarization figure of the electric field has $p_1+p_2$ separate leaves. The HH electric field can therefore be understood as an element of the cyclic group $C_{p_1+p_2}$ of order $p_1+p_2$. A multiple of a $\frac{2\pi}{p_1+p_2}$ rotation in the polarization plane leaves the polarization figure of the driving beam invariant. This invariance implies that the HHs need to preserve the symmetry as well. Therefore, the HH maxima mimic the polarization orientation of the Lissajous figure induced by the driving beam \cite{PhysRevA.62.011403}.

Since the polarization orientation of the HHs mimics the temporal and azimuthal evolution of the driving beam, the spatial torus knot can be constructed as illustrated in Fig.~\ref{setup}(f).
This qualitative analysis of the HHs yields an explanation for the conservation of TKAM  (and especially $\gamma$).

To underscore the above explanation, we present explicit numerical calculations in Fig.~\ref{azimuthalTorus} for the same beam parameters and harmonic orders that we used in Sec.~\ref{subsec:HHTemporalTorus}. The constant time-shift $\Delta t$ between the emission of two HH pulse maxima [Fig.~\ref{azimuthalTorus}(a)] demonstrates the association of each HH pulse with a certain intensity maximum of the driving beam. The explicit temporal evolution of the HHs is shown in Fig.~\ref{azimuthalTorus}(b). The thin gray lines display the respective projected contribution of the HHs and the color scheme denotes the increasing time. Here, the conservation of the polarization symmetry can be seen in the $E_x-E_y-$plane for $t=0$. Note that the polarization figure has six peaks instead of three since we do not take the modulus of the HH electric field. A comparison of Fig.~\ref{setup}(b) with Fig.~\ref{azimuthalTorus}(c) reveals the conservation of $\gamma$ and further agrees with the discussion above. The polarization figure of the HHs rotates by an angle of $2\pi\gamma=-\frac{2\pi}{3}$ on the target while the azimuthal angle $\varphi$ is increased from $0$ to $2\pi$. This can be seen explicitly in the rotation of the green dots of the polarization figure in Fig.~\ref{azimuthalTorus}(c) that mark a specific maximum of the HHs. 

\maketitle\subsection{\label{subsec:2Dmap}Geometric relation between spatial and temporal representations}
\begin{figure}[t!]
    \centering
    \includegraphics[scale=.33]{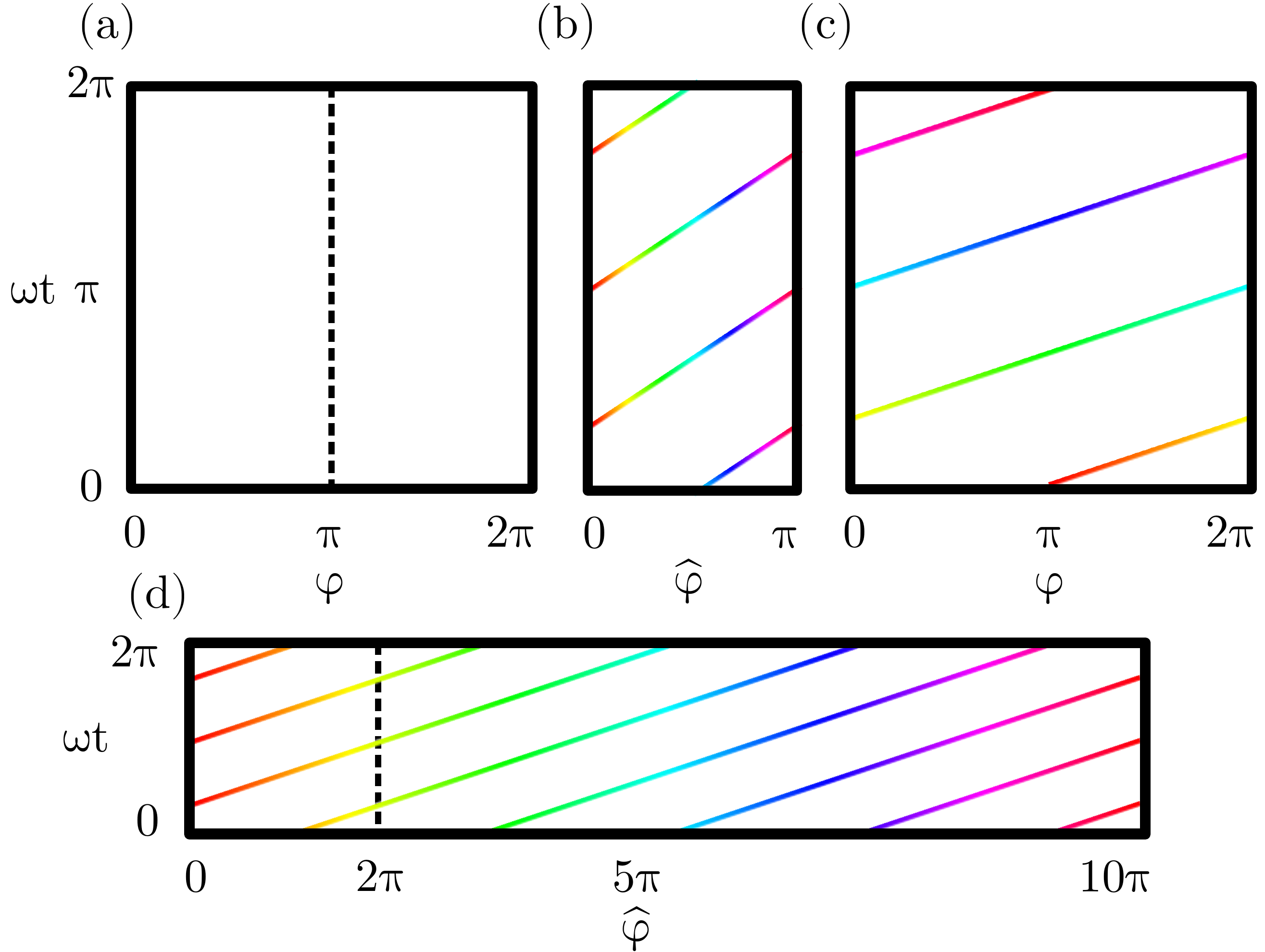}
    \caption{Schematic two-dimensional representation of the electric field maxima of the superimposed HHs generated from electric field maxima of the driving beams with (a), (b) $\ell_1=\ell_2=1,\;\omega_1=\omega_2/2=\omega$ and (c), (d) $\ell_1=-2,\;\ell_2=-1,\;\omega_1=\omega_2/2=\omega$. (a) and (c) are associated with the temporal torus knot (light spring) and correspond to the parameters $\tau=2/3,\gamma=-1/3$ and $\hat{\varphi}_{\mathrm{max}}=\pi$  while (b) and (d) are associated with the spatial torus knot and the corresponding parameters $\tau=-1/3,\gamma=5/3$ and $\hat{\varphi}_{\mathrm{max}}=10\pi$. The figures (b) and (c) are obtained by keeping only the respective left sides of the dashed line in figures (a) and (d). In order to guide the eye, equal colors represent equal toroidal coordinates of the resulting torus knot. Therefore, the dimension with evolving color is associated with the toroidal axis. The remaining poloidal dimension is bent to form a cylinder. Afterward, the respective open ends of the cylinder are connected to form a torus. The lines of the electric field maxima on the torus are the torus knots.}
    \label{2Dmap}
\end{figure}
In Secs.~\ref{subsec:HHTemporalTorus} and \ref{subsec:HHSpatialTorus} the invariance parameters $\tau$ and $\gamma$ of the TKAM operator were associated with the HHs.
Both derivations are based on the maxima of the HHs and their local polarization. In particular, the derivation is focused on the temporal and azimuthal evolution of the HH maxima. As discussed above, the HHG process conserves the symmetric structure of the driving beam. Therefore, the following investigations are valid not only for the driving beam but also for the HHs. The evolution of the electric field maxima can be represented as a pattern of tilted straight lines as illustrated in Fig.~\ref{2Dmap}. This two-dimensional representation of the electric field maxima classifies the time delay parameter $\tau$, see Fig.~\ref{2Dmap}(a) and (c), as well as the coordination parameter $\gamma$, see Fig.~\ref{2Dmap}(b) and (d). Furthermore, these two-dimensional representations can be associated either with a temporal or a spatial torus knot.

The specific difference between the spatial torus knot and the temporal torus knot is the order of operations performed to construct them and also a linear scaling factor with respect to the polarization rotation. The linear scaling factor is investigated in the following.
The time shift between two field maxima of the HHs is given by
\begin{align}
    \Delta_{t_0}=\frac{2\pi}{\omega_1+\omega_2}=\frac{T}{p_1+p_2}.
\end{align}
If we use If we use the symmetry transformation (\ref{CRsSym}) and the angle of rotation (\ref{DelayHHG}), the time shift can be transformed into the azimuthal shift $\Delta_{\varphi_0}=\frac{\Delta_{t_0}}{\tau}=\frac{2\pi}{\ell_1+\ell_2}$. The explicit values for $\ell_1=\ell_2=1,\;\omega_1=\omega_2/2=\omega$ are then $\Delta_{\varphi_0}=\pi$ and $\Delta_{t_0}=\frac{T}{3}$. These values can be recognized in Fig.~\ref{LSgraphic} with (c) for time $t=0$ and (f) for time $t=\frac{T}{3}$. In (c) the colored dots denote the explicit electric field maxima for azimuthal angles $\varphi=0,\pi$ and at time $t=0$. Here, we can see that the colored dotes are separated by $\Delta_{\varphi_0}=\pi$. In addition, the time shift is found by the comparison of the colored dots on the dashed Lissajous figure in Fig.~\ref{LSgraphic}(c) and (f).

Inserting $\Delta_{\varphi_0}$ into Eq.~(\ref{CRsSym}) leads to the specific rotation angle $\frac{2\pi}{\ell_1+\ell_2}\gamma$ of the Lissajous figure. In Fig.~\ref{LSgraphic}(c)-(f) this rotation angle is $\Delta_{\varphi_0}\gamma=-\frac{\pi}{3}$ and therefore the symmetry transformation reads
\begin{align}
    \bm{R}\left(\Delta_{\varphi_0}\gamma\right)\bm{E}\left(\bm{R}^{-1}\left(\Delta_{\varphi_0}\right)\bm{r},t\right)=\bm{E}\left(\bm{r},t+\Delta_{t_0}\right).
\end{align}
However, one needs to carefully consider also the boundary conditions of the spatial torus knot. To do so, we use the periodicity of the electric field with regard to the fundamental frequency $T$. Hence, the time shift and the associated azimuthal shift are given by $\Delta_t=(p_1+p_2)\Delta_{t_0}$ and $\Delta_{\varphi}=(p_1+p_2)\Delta_{\varphi_0}$, respectively. The polarization rotation concerning the temporal shift can be interpreted within a simple geometric context. The angle $\Delta_{\varphi}\gamma$ describes the rotation of the Lissajous figure induced by increasing (or decreasing) the azimuthal angle on the target to $\Delta_{\varphi}$. Here, we start from an electric field maximum at $\varphi_1=0$ until we reach the $(p_1+p_2)$\textit{th} maximum at $\varphi_{p_1+p_2}$. Note that the time remains fixed while increasing the azimuthal angle, see Fig.~\ref{LSgraphic}(c). Therefore, a translation in time from $t_0$ to $t_0+T$ induces a polarization rotation of $T\frac{\gamma}{\tau}$. This evolution of the Lissajous figure can be seen in Fig.~\ref{setup}(d).

We are now able to relate the temporal as well as the spatial torus knot to the two-dimensional representation of the field maxima in Fig.~\ref{2Dmap}. The azimuthal upper boundary for the temporal torus knot reads $\varphi_{max}=2\pi$ , while $\hat{\varphi}_{max}$ it is given by 
\begin{align}
    \hat{\varphi}_{\mathrm{max}}=|\frac{2\pi}{\omega}\frac{\ell_2p_1-\ell_1p_2}{\ell_1+\ell_2}|=|T\frac{\gamma}{\tau}|\label{upperLimit}
\end{align}
for the spatial torus knot. This change of the boundary is necessary since the inherent rotation of the electric field vector and the coordinate system need to be considered as well. In Fig.~\ref{2Dmap}, $\tau$ indicates the gradient with respect to the angles $\varphi$ and $\hat{\varphi}$, and $T=2\pi/\omega$ denotes the upper boundary of $t$. These relations provide the explanation for the upper boundary of $\hat{\varphi}$ in Eq.~(\ref{upperLimit})
\begin{align}
    \mathrm{max}\left[E_{H}(\hat{\varphi}(t))\right]=\tau\hat{\varphi}(t)=t\gamma,
\end{align}
where $\mathrm{max}\left[E_{H}\right]$ denotes the electric field maxima of the HHs. Here, we can see explicitly that the electric field maxima rotate $\gamma$ times in the polarization plane for $t\in\left[ t_0,t_0+T\right]$. This results in generalization of the approach of Pisanty \cite{Pisanty2019,Pisanty2019a} in order to relate the polarization rotation with regard to a $2\pi$ azimuthal rotation to the coordination parameter $\gamma$.  

Figure \ref{2Dmap} shows two examples of these two-dimensional representations. To explicitly form a temporal or spatial torus knot, respectively, the dimension with constant color need to be bent and connected. Note that, in Fig.~\ref{2Dmap}(c) no dimension with constant color seems to exist. Here we see that the line changes the color with time and therefore cannot be constant. Thus, in this example, the azimuthal dimension needs to be bent to form a cylinder. The remaining open ends of the cylinder need to be connected as well to form the respective torus knot.
These operations are the geometric representation of periodic boundary conditions.
In Figs.~\ref{2Dmap}(a) and (c) the associated boundary conditions are related to the time and the azimuthal angle. On the other hand, the boundary conditions associated with (b) and (d) are identified with the time and the \textit{rotation angle in the polarization plane}.

Bending and connecting the azimuthal dimension of the flat two-dimensional representation in Figs.~\ref{2Dmap}(a) and (c) leads to a cylinder which represents the light spring. The connection of both open ends yields the temporal torus knot that determines the time delay parameter $\tau$.

If we consider Figs.~\ref{2Dmap}(b) and (d) and interchange the order of connection we obtain the spatial torus knot which determines the coordination parameter $\gamma$. 
Further, spatial and temporal torus knots are classified within their evolution based on the azimuthal angle and time. Therefore, it is not surprising that the invariance parameters $\tau$ and $\gamma$ are determined by the factors $\omega_i$ and $\ell_i$ which appear in front of these respective dimensions ($\omega_it$, $\ell_i\varphi$) in Eq.~(\ref{electricfield}). 

The geometric approach pursued in this work provides an explicit explanation for the dependence of $\tau$ and $\gamma$ with regard to the beam parameters $\omega_i$ and $\ell_i$. The geometric approach demonstrated here classify TKAM is in good agreement with previous findings \cite{Ballantine2016,Pisanty2019,Pisanty2019a}. In addition, our approach provides a vivid model to illustrate and intuitively understand TKAM.
\section{\label{sec:Conclusion}Conclusion}
High harmonic generation with bicircular twisted light beams is an improving field in recent years. Therefore, a proper theoretical description of this process and the associated models is necessary. 
In this work, we developed a geometric method to determine the time delay parameter $\tau$ associated with the TKAM in the context of HHG of planar atomic gas targets.

We showed that the time delay parameter $\tau$ can be read off of the temporal evolution of the intensity distribution of the high harmonic radiation. This temporal evolution yields a spiral structure of the electric field of the emitted radiation, also called a \textit{light spring}. The method developed here allowed us to determine the TKAM from the high harmonic radiation, as well as from the driving beam.

In addition, we demonstrated that the polarization rotation of the high harmonic radiation mirrors the one of the driving beam. Therefore, we explicitly presented a model which explains the conservation of TKAM and especially the conservation of the so-called coordination parameter $\gamma$.

Finally, we calculated and visualized the geometric relation between $\tau$ and $\gamma$. Both parameters $\tau$ and $\gamma$ are associated with torus knots. The underlying tori can be constructed from the same initial high harmonic field with a minor adaption in the boundary conditions. Therefore, the high harmonic radiation, as well as the driving beam, can be reduced to their temporal and azimuthal angle-dependent electric field maxima. The evolution of these maxima determines the TKAM and further forms a \textit{spatial torus knot} or a \textit{temporal torus knot} associated with $\tau$ or $\gamma$, respectively.

Investigations in the future may consider three-dimensional targets and finite driving pulse durations of the target to analyze the symmetry properties of TKAM in specific experimental setups. Furthermore, interesting properties of the TKAM may be found while increasing the harmonic order of the HHs beyond the classical cutoff region or by investigating LG beams with radial node number $p\neq0$.   
\section*{\label{sec:Acknowledgement}Acknowledgement}
This work has been funded by the Deutsche Forschungsgemeinschaft (DFG, German Research Foundation) under project number 440556973. The authors thank Willi Paufler for helpful conversations and and assistence with the 3D graphics.
\bibliography{references} 

\providecommand{\noopsort}[1]{}\providecommand{\singleletter}[1]{#1}%
\begin{thebibliography}{35}%
\makeatletter
\providecommand \@ifxundefined [1]{%
 \@ifx{#1\undefined}
}%
\providecommand \@ifnum [1]{%
 \ifnum #1\expandafter \@firstoftwo
 \else \expandafter \@secondoftwo
 \fi
}%
\providecommand \@ifx [1]{%
 \ifx #1\expandafter \@firstoftwo
 \else \expandafter \@secondoftwo
 \fi
}%
\providecommand \natexlab [1]{#1}%
\providecommand \enquote  [1]{``#1''}%
\providecommand \bibnamefont  [1]{#1}%
\providecommand \bibfnamefont [1]{#1}%
\providecommand \citenamefont [1]{#1}%
\providecommand \href@noop [0]{\@secondoftwo}%
\providecommand \href [0]{\begingroup \@sanitize@url \@href}%
\providecommand \@href[1]{\@@startlink{#1}\@@href}%
\providecommand \@@href[1]{\endgroup#1\@@endlink}%
\providecommand \@sanitize@url [0]{\catcode `\\12\catcode `\$12\catcode
  `\&12\catcode `\#12\catcode `\^12\catcode `\_12\catcode `\%12\relax}%
\providecommand \@@startlink[1]{}%
\providecommand \@@endlink[0]{}%
\providecommand \url  [0]{\begingroup\@sanitize@url \@url }%
\providecommand \@url [1]{\endgroup\@href {#1}{\urlprefix }}%
\providecommand \urlprefix  [0]{URL }%
\providecommand \Eprint [0]{\href }%
\providecommand \doibase [0]{https://doi.org/}%
\providecommand \selectlanguage [0]{\@gobble}%
\providecommand \bibinfo  [0]{\@secondoftwo}%
\providecommand \bibfield  [0]{\@secondoftwo}%
\providecommand \translation [1]{[#1]}%
\providecommand \BibitemOpen [0]{}%
\providecommand \bibitemStop [0]{}%
\providecommand \bibitemNoStop [0]{.\EOS\space}%
\providecommand \EOS [0]{\spacefactor3000\relax}%
\providecommand \BibitemShut  [1]{\csname bibitem#1\endcsname}%
\let\auto@bib@innerbib\@empty
\bibitem [{\citenamefont {Agostini}\ \emph {et~al.}(1979)\citenamefont
  {Agostini}, \citenamefont {Fabre}, \citenamefont {Mainfray}, \citenamefont
  {Petite},\ and\ \citenamefont {Rahman}}]{Agostini1979}%
  \BibitemOpen
  \bibfield  {author} {\bibinfo {author} {\bibfnamefont {P.}~\bibnamefont
  {Agostini}}, \bibinfo {author} {\bibfnamefont {F.}~\bibnamefont {Fabre}},
  \bibinfo {author} {\bibfnamefont {G.}~\bibnamefont {Mainfray}}, \bibinfo
  {author} {\bibfnamefont {G.}~\bibnamefont {Petite}},\ and\ \bibinfo {author}
  {\bibfnamefont {N.~K.}\ \bibnamefont {Rahman}},\ }\bibfield  {title}
  {\bibinfo {title} {Free-free transitions following six-photon ionization of
  xenon atoms},\ }\href {https://doi.org/10.1103/PhysRevLett.42.1127}
  {\bibfield  {journal} {\bibinfo  {journal} {Phys. Rev. Lett.}\ }\textbf
  {\bibinfo {volume} {42}},\ \bibinfo {pages} {1127} (\bibinfo {year}
  {1979})}\BibitemShut {NoStop}%
\bibitem [{\citenamefont {Paulus}\ \emph {et~al.}(1994)\citenamefont {Paulus},
  \citenamefont {Nicklich}, \citenamefont {Xu}, \citenamefont {Lambropoulos},\
  and\ \citenamefont {Walther}}]{Paulus1994}%
  \BibitemOpen
  \bibfield  {author} {\bibinfo {author} {\bibfnamefont {G.~G.}\ \bibnamefont
  {Paulus}}, \bibinfo {author} {\bibfnamefont {W.}~\bibnamefont {Nicklich}},
  \bibinfo {author} {\bibfnamefont {H.}~\bibnamefont {Xu}}, \bibinfo {author}
  {\bibfnamefont {P.}~\bibnamefont {Lambropoulos}},\ and\ \bibinfo {author}
  {\bibfnamefont {H.}~\bibnamefont {Walther}},\ }\bibfield  {title} {\bibinfo
  {title} {Plateau in above threshold ionization spectra},\ }\href
  {https://doi.org/10.1103/PhysRevLett.72.2851} {\bibfield  {journal} {\bibinfo
   {journal} {Phys. Rev. Lett.}\ }\textbf {\bibinfo {volume} {72}},\ \bibinfo
  {pages} {2851} (\bibinfo {year} {1994})}\BibitemShut {NoStop}%
\bibitem [{\citenamefont {McPherson}\ \emph {et~al.}(1987)\citenamefont
  {McPherson}, \citenamefont {Gibson}, \citenamefont {Jara}, \citenamefont
  {Johann}, \citenamefont {Luk}, \citenamefont {McIntyre}, \citenamefont
  {Boyer},\ and\ \citenamefont {Rhodes}}]{McPherson1987}%
  \BibitemOpen
  \bibfield  {author} {\bibinfo {author} {\bibfnamefont {A.}~\bibnamefont
  {McPherson}}, \bibinfo {author} {\bibfnamefont {G.}~\bibnamefont {Gibson}},
  \bibinfo {author} {\bibfnamefont {H.}~\bibnamefont {Jara}}, \bibinfo {author}
  {\bibfnamefont {U.}~\bibnamefont {Johann}}, \bibinfo {author} {\bibfnamefont
  {T.~S.}\ \bibnamefont {Luk}}, \bibinfo {author} {\bibfnamefont {I.~A.}\
  \bibnamefont {McIntyre}}, \bibinfo {author} {\bibfnamefont {K.}~\bibnamefont
  {Boyer}},\ and\ \bibinfo {author} {\bibfnamefont {C.~K.}\ \bibnamefont
  {Rhodes}},\ }\bibfield  {title} {\bibinfo {title} {Studies of multiphoton
  production of vacuum-ultraviolet radiation in the rare gases},\ }\href
  {https://doi.org/10.1364/JOSAB.4.000595} {\bibfield  {journal} {\bibinfo
  {journal} {J. Opt. Soc. Am. B}\ }\textbf {\bibinfo {volume} {4}},\ \bibinfo
  {pages} {595} (\bibinfo {year} {1987})}\BibitemShut {NoStop}%
\bibitem [{\citenamefont {Ferray}\ \emph {et~al.}(1988)\citenamefont {Ferray},
  \citenamefont {L{\textquotesingle}Huillier}, \citenamefont {Li},
  \citenamefont {Lompre}, \citenamefont {Mainfray},\ and\ \citenamefont
  {Manus}}]{Ferray1988}%
  \BibitemOpen
  \bibfield  {author} {\bibinfo {author} {\bibfnamefont {M.}~\bibnamefont
  {Ferray}}, \bibinfo {author} {\bibfnamefont {A.}~\bibnamefont
  {L{\textquotesingle}Huillier}}, \bibinfo {author} {\bibfnamefont {X.~F.}\
  \bibnamefont {Li}}, \bibinfo {author} {\bibfnamefont {L.~A.}\ \bibnamefont
  {Lompre}}, \bibinfo {author} {\bibfnamefont {G.}~\bibnamefont {Mainfray}},\
  and\ \bibinfo {author} {\bibfnamefont {C.}~\bibnamefont {Manus}},\ }\bibfield
   {title} {\bibinfo {title} {Multiple-harmonic conversion of 1064 nm radiation
  in rare gases},\ }\href {https://doi.org/10.1088/0953-4075/21/3/001}
  {\bibfield  {journal} {\bibinfo  {journal} {Journal of Physics B: Atomic,
  Molecular and Optical Physics}\ }\textbf {\bibinfo {volume} {21}},\ \bibinfo
  {pages} {L31} (\bibinfo {year} {1988})}\BibitemShut {NoStop}%
\bibitem [{\citenamefont {Frolov}\ \emph {et~al.}(2018)\citenamefont {Frolov},
  \citenamefont {Manakov}, \citenamefont {Minina}, \citenamefont {Vvedenskii},
  \citenamefont {Silaev}, \citenamefont {Ivanov},\ and\ \citenamefont
  {Starace}}]{Frolov2018}%
  \BibitemOpen
  \bibfield  {author} {\bibinfo {author} {\bibfnamefont {M.~V.}\ \bibnamefont
  {Frolov}}, \bibinfo {author} {\bibfnamefont {N.~L.}\ \bibnamefont {Manakov}},
  \bibinfo {author} {\bibfnamefont {A.~A.}\ \bibnamefont {Minina}}, \bibinfo
  {author} {\bibfnamefont {N.~V.}\ \bibnamefont {Vvedenskii}}, \bibinfo
  {author} {\bibfnamefont {A.~A.}\ \bibnamefont {Silaev}}, \bibinfo {author}
  {\bibfnamefont {M.~Y.}\ \bibnamefont {Ivanov}},\ and\ \bibinfo {author}
  {\bibfnamefont {A.~F.}\ \bibnamefont {Starace}},\ }\bibfield  {title}
  {\bibinfo {title} {Control of harmonic generation by the time delay between
  two-color, bicircular few-cycle mid-ir laser pulses},\ }\href
  {https://doi.org/10.1103/PhysRevLett.120.263203} {\bibfield  {journal}
  {\bibinfo  {journal} {Phys. Rev. Lett.}\ }\textbf {\bibinfo {volume} {120}},\
  \bibinfo {pages} {263203} (\bibinfo {year} {2018})}\BibitemShut {NoStop}%
\bibitem [{\citenamefont {Milo\ifmmode \check{s}\else
  \v{s}\fi{}evi\ifmmode~\acute{c}\else \'{c}\fi{}}(2018)}]{Milo2018}%
  \BibitemOpen
  \bibfield  {author} {\bibinfo {author} {\bibfnamefont {D.~B.}\ \bibnamefont
  {Milo\ifmmode \check{s}\else \v{s}\fi{}evi\ifmmode~\acute{c}\else
  \'{c}\fi{}}},\ }\bibfield  {title} {\bibinfo {title} {Low-frequency
  approximation for high-order harmonic generation by a bicircular laser
  field},\ }\href {https://doi.org/10.1103/PhysRevA.97.013416} {\bibfield
  {journal} {\bibinfo  {journal} {Phys. Rev. A}\ }\textbf {\bibinfo {volume}
  {97}},\ \bibinfo {pages} {013416} (\bibinfo {year} {2018})}\BibitemShut
  {NoStop}%
\bibitem [{\citenamefont {Rego}\ \emph {et~al.}(2019)\citenamefont {Rego},
  \citenamefont {Dorney}, \citenamefont {Brooks}, \citenamefont {Nguyen},
  \citenamefont {Liao}, \citenamefont {Rom{\'{a}}n}, \citenamefont {Couch},
  \citenamefont {Liu}, \citenamefont {Pisanty}, \citenamefont {Lewenstein},
  \citenamefont {Plaja}, \citenamefont {Kapteyn}, \citenamefont {Murnane},\
  and\ \citenamefont {Hern{\'{a}}ndez-Garc{\'{\i}}a}}]{Rego2019}%
  \BibitemOpen
  \bibfield  {author} {\bibinfo {author} {\bibfnamefont {L.}~\bibnamefont
  {Rego}}, \bibinfo {author} {\bibfnamefont {K.~M.}\ \bibnamefont {Dorney}},
  \bibinfo {author} {\bibfnamefont {N.~J.}\ \bibnamefont {Brooks}}, \bibinfo
  {author} {\bibfnamefont {Q.~L.}\ \bibnamefont {Nguyen}}, \bibinfo {author}
  {\bibfnamefont {C.-T.}\ \bibnamefont {Liao}}, \bibinfo {author}
  {\bibfnamefont {J.~S.}\ \bibnamefont {Rom{\'{a}}n}}, \bibinfo {author}
  {\bibfnamefont {D.~E.}\ \bibnamefont {Couch}}, \bibinfo {author}
  {\bibfnamefont {A.}~\bibnamefont {Liu}}, \bibinfo {author} {\bibfnamefont
  {E.}~\bibnamefont {Pisanty}}, \bibinfo {author} {\bibfnamefont
  {M.}~\bibnamefont {Lewenstein}}, \bibinfo {author} {\bibfnamefont
  {L.}~\bibnamefont {Plaja}}, \bibinfo {author} {\bibfnamefont {H.~C.}\
  \bibnamefont {Kapteyn}}, \bibinfo {author} {\bibfnamefont {M.~M.}\
  \bibnamefont {Murnane}},\ and\ \bibinfo {author} {\bibfnamefont
  {C.}~\bibnamefont {Hern{\'{a}}ndez-Garc{\'{\i}}a}},\ }\bibfield  {title}
  {\bibinfo {title} {Generation of extreme-ultraviolet beams with time-varying
  orbital angular momentum},\ }\href {https://doi.org/10.1126/science.aaw9486}
  {\bibfield  {journal} {\bibinfo  {journal} {Science}\ }\textbf {\bibinfo
  {volume} {364}},\ \bibinfo {pages} {eaaw9486} (\bibinfo {year}
  {2019})}\BibitemShut {NoStop}%
\bibitem [{\citenamefont {Dorney}\ \emph {et~al.}(2018)\citenamefont {Dorney},
  \citenamefont {Rego}, \citenamefont {Brooks}, \citenamefont {Rom{\'{a}}n},
  \citenamefont {Liao}, \citenamefont {Ellis}, \citenamefont {Zusin},
  \citenamefont {Gentry}, \citenamefont {Nguyen}, \citenamefont {Shaw},
  \citenamefont {Pic{\'{o}}n}, \citenamefont {Plaja}, \citenamefont {Kapteyn},
  \citenamefont {Murnane},\ and\ \citenamefont
  {Hern{\'{a}}ndez-Garc{\'{\i}}a}}]{Dorney2018}%
  \BibitemOpen
  \bibfield  {author} {\bibinfo {author} {\bibfnamefont {K.~M.}\ \bibnamefont
  {Dorney}}, \bibinfo {author} {\bibfnamefont {L.}~\bibnamefont {Rego}},
  \bibinfo {author} {\bibfnamefont {N.~J.}\ \bibnamefont {Brooks}}, \bibinfo
  {author} {\bibfnamefont {J.~S.}\ \bibnamefont {Rom{\'{a}}n}}, \bibinfo
  {author} {\bibfnamefont {C.-T.}\ \bibnamefont {Liao}}, \bibinfo {author}
  {\bibfnamefont {J.~L.}\ \bibnamefont {Ellis}}, \bibinfo {author}
  {\bibfnamefont {D.}~\bibnamefont {Zusin}}, \bibinfo {author} {\bibfnamefont
  {C.}~\bibnamefont {Gentry}}, \bibinfo {author} {\bibfnamefont {Q.~L.}\
  \bibnamefont {Nguyen}}, \bibinfo {author} {\bibfnamefont {J.~M.}\
  \bibnamefont {Shaw}}, \bibinfo {author} {\bibfnamefont {A.}~\bibnamefont
  {Pic{\'{o}}n}}, \bibinfo {author} {\bibfnamefont {L.}~\bibnamefont {Plaja}},
  \bibinfo {author} {\bibfnamefont {H.~C.}\ \bibnamefont {Kapteyn}}, \bibinfo
  {author} {\bibfnamefont {M.~M.}\ \bibnamefont {Murnane}},\ and\ \bibinfo
  {author} {\bibfnamefont {C.}~\bibnamefont {Hern{\'{a}}ndez-Garc{\'{\i}}a}},\
  }\bibfield  {title} {\bibinfo {title} {Controlling the polarization and
  vortex charge of attosecond high-harmonic beams via simultaneous
  spin{\textendash}orbit momentum conservation},\ }\href
  {https://doi.org/10.1038/s41566-018-0304-3} {\bibfield  {journal} {\bibinfo
  {journal} {Nature Photonics}\ }\textbf {\bibinfo {volume} {13}},\ \bibinfo
  {pages} {123} (\bibinfo {year} {2018})}\BibitemShut {NoStop}%
\bibitem [{\citenamefont {Paufler}\ \emph {et~al.}(2018)\citenamefont
  {Paufler}, \citenamefont {B\"oning},\ and\ \citenamefont
  {Fritzsche}}]{Paufler2018}%
  \BibitemOpen
  \bibfield  {author} {\bibinfo {author} {\bibfnamefont {W.}~\bibnamefont
  {Paufler}}, \bibinfo {author} {\bibfnamefont {B.}~\bibnamefont {B\"oning}},\
  and\ \bibinfo {author} {\bibfnamefont {S.}~\bibnamefont {Fritzsche}},\
  }\bibfield  {title} {\bibinfo {title} {Tailored orbital angular momentum in
  high-order harmonic generation with bicircular laguerre-gaussian beams},\
  }\href {https://doi.org/10.1103/PhysRevA.98.011401} {\bibfield  {journal}
  {\bibinfo  {journal} {Phys. Rev. A}\ }\textbf {\bibinfo {volume} {98}},\
  \bibinfo {pages} {011401} (\bibinfo {year} {2018})}\BibitemShut {NoStop}%
\bibitem [{\citenamefont {Paufler}\ \emph {et~al.}(2019)\citenamefont
  {Paufler}, \citenamefont {Böning},\ and\ \citenamefont
  {Fritzsche}}]{Paufler2019}%
  \BibitemOpen
  \bibfield  {author} {\bibinfo {author} {\bibfnamefont {W.}~\bibnamefont
  {Paufler}}, \bibinfo {author} {\bibfnamefont {B.}~\bibnamefont {Böning}},\
  and\ \bibinfo {author} {\bibfnamefont {S.}~\bibnamefont {Fritzsche}},\
  }\bibfield  {title} {\bibinfo {title} {High harmonic generation with
  laguerre{\textendash}gaussian beams},\ }\href
  {https://doi.org/10.1088/2040-8986/ab31c3} {\bibfield  {journal} {\bibinfo
  {journal} {Journal of Optics}\ }\textbf {\bibinfo {volume} {21}},\ \bibinfo
  {pages} {094001} (\bibinfo {year} {2019})}\BibitemShut {NoStop}%
\bibitem [{\citenamefont {Hernández-García}\ \emph
  {et~al.}(2017)\citenamefont {Hernández-García}, \citenamefont {Rego},
  \citenamefont {San~Román}, \citenamefont {Picón},\ and\ \citenamefont
  {Plaja}}]{Garcia2017}%
  \BibitemOpen
  \bibfield  {author} {\bibinfo {author} {\bibfnamefont {C.}~\bibnamefont
  {Hernández-García}}, \bibinfo {author} {\bibfnamefont {L.}~\bibnamefont
  {Rego}}, \bibinfo {author} {\bibfnamefont {J.}~\bibnamefont {San~Román}},
  \bibinfo {author} {\bibfnamefont {A.}~\bibnamefont {Picón}},\ and\ \bibinfo
  {author} {\bibfnamefont {L.}~\bibnamefont {Plaja}},\ }\bibfield  {title}
  {\bibinfo {title} {Attosecond twisted beams from high-order harmonic
  generation driven by optical vortices},\ }\href
  {https://doi.org/10.1017/hpl.2017.1} {\bibfield  {journal} {\bibinfo
  {journal} {High Power Laser Science and Engineering}\ }\textbf {\bibinfo
  {volume} {5}},\ \bibinfo {pages} {e3} (\bibinfo {year} {2017})}\BibitemShut
  {NoStop}%
\bibitem [{\citenamefont {Gauthier}\ \emph {et~al.}(2017)\citenamefont
  {Gauthier}, \citenamefont {Ribi{\v{c}}}, \citenamefont {Adhikary},
  \citenamefont {Camper}, \citenamefont {Chappuis}, \citenamefont {Cucini},
  \citenamefont {DiMauro}, \citenamefont {Dovillaire}, \citenamefont
  {Frassetto}, \citenamefont {G{\'e}neaux} \emph {et~al.}}]{Gauthier2017}%
  \BibitemOpen
  \bibfield  {author} {\bibinfo {author} {\bibfnamefont {D.}~\bibnamefont
  {Gauthier}}, \bibinfo {author} {\bibfnamefont {P.~R.}\ \bibnamefont
  {Ribi{\v{c}}}}, \bibinfo {author} {\bibfnamefont {G.}~\bibnamefont
  {Adhikary}}, \bibinfo {author} {\bibfnamefont {A.}~\bibnamefont {Camper}},
  \bibinfo {author} {\bibfnamefont {C.}~\bibnamefont {Chappuis}}, \bibinfo
  {author} {\bibfnamefont {R.}~\bibnamefont {Cucini}}, \bibinfo {author}
  {\bibfnamefont {L.}~\bibnamefont {DiMauro}}, \bibinfo {author} {\bibfnamefont
  {G.}~\bibnamefont {Dovillaire}}, \bibinfo {author} {\bibfnamefont
  {F.}~\bibnamefont {Frassetto}}, \bibinfo {author} {\bibfnamefont
  {R.}~\bibnamefont {G{\'e}neaux}}, \emph {et~al.},\ }\bibfield  {title}
  {\bibinfo {title} {Tunable orbital angular momentum in high-harmonic
  generation},\ }\href@noop {} {\bibfield  {journal} {\bibinfo  {journal}
  {Nature communications}\ }\textbf {\bibinfo {volume} {8}},\ \bibinfo {pages}
  {1} (\bibinfo {year} {2017})}\BibitemShut {NoStop}%
\bibitem [{\citenamefont {Devlin}\ \emph {et~al.}(2017)\citenamefont {Devlin},
  \citenamefont {Ambrosio}, \citenamefont {Rubin}, \citenamefont {Mueller},\
  and\ \citenamefont {Capasso}}]{Devlin2017}%
  \BibitemOpen
  \bibfield  {author} {\bibinfo {author} {\bibfnamefont {R.~C.}\ \bibnamefont
  {Devlin}}, \bibinfo {author} {\bibfnamefont {A.}~\bibnamefont {Ambrosio}},
  \bibinfo {author} {\bibfnamefont {N.~A.}\ \bibnamefont {Rubin}}, \bibinfo
  {author} {\bibfnamefont {J.~P.~B.}\ \bibnamefont {Mueller}},\ and\ \bibinfo
  {author} {\bibfnamefont {F.}~\bibnamefont {Capasso}},\ }\bibfield  {title}
  {\bibinfo {title} {Arbitrary spin-to{\textendash}orbital angular momentum
  conversion of light},\ }\href {https://doi.org/10.1126/science.aao5392}
  {\bibfield  {journal} {\bibinfo  {journal} {Science}\ }\textbf {\bibinfo
  {volume} {358}},\ \bibinfo {pages} {896} (\bibinfo {year} {2017})},\ \Eprint
  {https://arxiv.org/abs/https://science.sciencemag.org/content/358/6365/896.full.pdf}
  {https://science.sciencemag.org/content/358/6365/896.full.pdf} \BibitemShut
  {NoStop}%
\bibitem [{\citenamefont {Ballantine}\ \emph {et~al.}(2016)\citenamefont
  {Ballantine}, \citenamefont {Donegan},\ and\ \citenamefont
  {Eastham}}]{Ballantine2016}%
  \BibitemOpen
  \bibfield  {author} {\bibinfo {author} {\bibfnamefont {K.~E.}\ \bibnamefont
  {Ballantine}}, \bibinfo {author} {\bibfnamefont {J.~F.}\ \bibnamefont
  {Donegan}},\ and\ \bibinfo {author} {\bibfnamefont {P.~R.}\ \bibnamefont
  {Eastham}},\ }\bibfield  {title} {\bibinfo {title} {{There are many ways to
  spin a photon: Half-quantization of a total optical angular momentum}},\
  }\bibfield  {journal} {\bibinfo  {journal} {Science Advances}\ }\textbf
  {\bibinfo {volume} {2}},\ \href {https://doi.org/10.1126/sciadv.1501748}
  {10.1126/sciadv.1501748} (\bibinfo {year} {2016})\BibitemShut {NoStop}%
\bibitem [{\citenamefont {Allen}\ \emph {et~al.}(1992)\citenamefont {Allen},
  \citenamefont {Beijersbergen}, \citenamefont {Spreeuw},\ and\ \citenamefont
  {Woerdman}}]{Allen1992}%
  \BibitemOpen
  \bibfield  {author} {\bibinfo {author} {\bibfnamefont {L.}~\bibnamefont
  {Allen}}, \bibinfo {author} {\bibfnamefont {M.~W.}\ \bibnamefont
  {Beijersbergen}}, \bibinfo {author} {\bibfnamefont {R.~J.~C.}\ \bibnamefont
  {Spreeuw}},\ and\ \bibinfo {author} {\bibfnamefont {J.~P.}\ \bibnamefont
  {Woerdman}},\ }\bibfield  {title} {\bibinfo {title} {Orbital angular momentum
  of light and the transformation of laguerre-gaussian laser modes},\ }\href
  {https://doi.org/10.1103/PhysRevA.45.8185} {\bibfield  {journal} {\bibinfo
  {journal} {Phys. Rev. A}\ }\textbf {\bibinfo {volume} {45}},\ \bibinfo
  {pages} {8185} (\bibinfo {year} {1992})}\BibitemShut {NoStop}%
\bibitem [{\citenamefont {Harris}\ \emph {et~al.}(1994)\citenamefont {Harris},
  \citenamefont {Hill}, \citenamefont {Tapster},\ and\ \citenamefont
  {Vaughan}}]{Harris1994}%
  \BibitemOpen
  \bibfield  {author} {\bibinfo {author} {\bibfnamefont {M.}~\bibnamefont
  {Harris}}, \bibinfo {author} {\bibfnamefont {C.~A.}\ \bibnamefont {Hill}},
  \bibinfo {author} {\bibfnamefont {P.~R.}\ \bibnamefont {Tapster}},\ and\
  \bibinfo {author} {\bibfnamefont {J.~M.}\ \bibnamefont {Vaughan}},\
  }\bibfield  {title} {\bibinfo {title} {Laser modes with helical wave
  fronts},\ }\href {https://doi.org/10.1103/PhysRevA.49.3119} {\bibfield
  {journal} {\bibinfo  {journal} {Phys. Rev. A}\ }\textbf {\bibinfo {volume}
  {49}},\ \bibinfo {pages} {3119} (\bibinfo {year} {1994})}\BibitemShut
  {NoStop}%
\bibitem [{\citenamefont {Bazhenov}\ \emph {et~al.}(1990)\citenamefont
  {Bazhenov}, \citenamefont {Vasnetsov},\ and\ \citenamefont
  {Soskin}}]{Bazhenov1990}%
  \BibitemOpen
  \bibfield  {author} {\bibinfo {author} {\bibfnamefont {V.~Y.}\ \bibnamefont
  {Bazhenov}}, \bibinfo {author} {\bibfnamefont {M.}~\bibnamefont
  {Vasnetsov}},\ and\ \bibinfo {author} {\bibfnamefont {M.}~\bibnamefont
  {Soskin}},\ }\bibfield  {title} {\bibinfo {title} {Laser beams with screw
  dislocations in their wavefronts},\ }\href@noop {} {\bibfield  {journal}
  {\bibinfo  {journal} {Jetp Lett}\ }\textbf {\bibinfo {volume} {52}},\
  \bibinfo {pages} {429} (\bibinfo {year} {1990})}\BibitemShut {NoStop}%
\bibitem [{\citenamefont {Pisanty}\ \emph
  {et~al.}(2019{\natexlab{a}})\citenamefont {Pisanty}, \citenamefont {Machado},
  \citenamefont {Vicu{\~{n}}a-Hern{\'{a}}ndez}, \citenamefont {Pic{\'{o}}n},
  \citenamefont {Celi}, \citenamefont {Torres},\ and\ \citenamefont
  {Lewenstein}}]{Pisanty2019a}%
  \BibitemOpen
  \bibfield  {author} {\bibinfo {author} {\bibfnamefont {E.}~\bibnamefont
  {Pisanty}}, \bibinfo {author} {\bibfnamefont {G.~J.}\ \bibnamefont
  {Machado}}, \bibinfo {author} {\bibfnamefont {V.}~\bibnamefont
  {Vicu{\~{n}}a-Hern{\'{a}}ndez}}, \bibinfo {author} {\bibfnamefont
  {A.}~\bibnamefont {Pic{\'{o}}n}}, \bibinfo {author} {\bibfnamefont
  {A.}~\bibnamefont {Celi}}, \bibinfo {author} {\bibfnamefont {J.~P.}\
  \bibnamefont {Torres}},\ and\ \bibinfo {author} {\bibfnamefont
  {M.}~\bibnamefont {Lewenstein}},\ }\bibfield  {title} {\bibinfo {title}
  {Knotting fractional-order knots with the polarization state of light},\
  }\href {https://doi.org/10.1038/s41566-019-0450-2} {\bibfield  {journal}
  {\bibinfo  {journal} {Nature Photonics}\ }\textbf {\bibinfo {volume} {13}},\
  \bibinfo {pages} {569} (\bibinfo {year} {2019}{\natexlab{a}})}\BibitemShut
  {NoStop}%
\bibitem [{\citenamefont {Pisanty}\ \emph
  {et~al.}(2019{\natexlab{b}})\citenamefont {Pisanty}, \citenamefont {Rego},
  \citenamefont {San~Rom\'an}, \citenamefont {Pic\'on}, \citenamefont {Dorney},
  \citenamefont {Kapteyn}, \citenamefont {Murnane}, \citenamefont {Plaja},
  \citenamefont {Lewenstein},\ and\ \citenamefont
  {Hern\'andez-Garc\'{\i}a}}]{Pisanty2019}%
  \BibitemOpen
  \bibfield  {author} {\bibinfo {author} {\bibfnamefont {E.}~\bibnamefont
  {Pisanty}}, \bibinfo {author} {\bibfnamefont {L.}~\bibnamefont {Rego}},
  \bibinfo {author} {\bibfnamefont {J.}~\bibnamefont {San~Rom\'an}}, \bibinfo
  {author} {\bibfnamefont {A.}~\bibnamefont {Pic\'on}}, \bibinfo {author}
  {\bibfnamefont {K.~M.}\ \bibnamefont {Dorney}}, \bibinfo {author}
  {\bibfnamefont {H.~C.}\ \bibnamefont {Kapteyn}}, \bibinfo {author}
  {\bibfnamefont {M.~M.}\ \bibnamefont {Murnane}}, \bibinfo {author}
  {\bibfnamefont {L.}~\bibnamefont {Plaja}}, \bibinfo {author} {\bibfnamefont
  {M.}~\bibnamefont {Lewenstein}},\ and\ \bibinfo {author} {\bibfnamefont
  {C.}~\bibnamefont {Hern\'andez-Garc\'{\i}a}},\ }\bibfield  {title} {\bibinfo
  {title} {Conservation of torus-knot angular momentum in high-order harmonic
  generation},\ }\href {https://doi.org/10.1103/PhysRevLett.122.203201}
  {\bibfield  {journal} {\bibinfo  {journal} {Phys. Rev. Lett.}\ }\textbf
  {\bibinfo {volume} {122}},\ \bibinfo {pages} {203201} (\bibinfo {year}
  {2019}{\natexlab{b}})}\BibitemShut {NoStop}%
\bibitem [{And(2012)}]{Andrews2012}%
  \BibitemOpen
  \href {https://doi.org/10.1017/CBO9780511795213} {\emph {\bibinfo {title}
  {The Angular Momentum of Light}}}\ (\bibinfo  {publisher} {Cambridge
  University Press},\ \bibinfo {year} {2012})\BibitemShut {NoStop}%
\bibitem [{\citenamefont {Bliokh}\ \emph {et~al.}(2015)\citenamefont {Bliokh},
  \citenamefont {Rodr{\'i}guez-Fortu{\~{n}}o}, \citenamefont {Nori},\ and\
  \citenamefont {Zayats}}]{Bliokh2015}%
  \BibitemOpen
  \bibfield  {author} {\bibinfo {author} {\bibfnamefont {K.~Y.}\ \bibnamefont
  {Bliokh}}, \bibinfo {author} {\bibfnamefont {F.~J.}\ \bibnamefont
  {Rodr{\'i}guez-Fortu{\~{n}}o}}, \bibinfo {author} {\bibfnamefont
  {F.}~\bibnamefont {Nori}},\ and\ \bibinfo {author} {\bibfnamefont {A.~V.}\
  \bibnamefont {Zayats}},\ }\bibfield  {title} {\bibinfo {title} {Spin--orbit
  interactions of light},\ }\href {https://doi.org/10.1038/nphoton.2015.201}
  {\bibfield  {journal} {\bibinfo  {journal} {Nature Photonics}\ }\textbf
  {\bibinfo {volume} {9}},\ \bibinfo {pages} {796} (\bibinfo {year}
  {2015})}\BibitemShut {NoStop}%
\bibitem [{\citenamefont {Lewenstein}\ \emph {et~al.}(1994)\citenamefont
  {Lewenstein}, \citenamefont {Balcou}, \citenamefont {Ivanov}, \citenamefont
  {L'Huillier},\ and\ \citenamefont {Corkum}}]{Lewenstein1994}%
  \BibitemOpen
  \bibfield  {author} {\bibinfo {author} {\bibfnamefont {M.}~\bibnamefont
  {Lewenstein}}, \bibinfo {author} {\bibfnamefont {P.}~\bibnamefont {Balcou}},
  \bibinfo {author} {\bibfnamefont {M.~Y.}\ \bibnamefont {Ivanov}}, \bibinfo
  {author} {\bibfnamefont {A.}~\bibnamefont {L'Huillier}},\ and\ \bibinfo
  {author} {\bibfnamefont {P.~B.}\ \bibnamefont {Corkum}},\ }\bibfield  {title}
  {\bibinfo {title} {Theory of high-harmonic generation by low-frequency laser
  fields},\ }\href {https://doi.org/10.1103/PhysRevA.49.2117} {\bibfield
  {journal} {\bibinfo  {journal} {Phys. Rev. A}\ }\textbf {\bibinfo {volume}
  {49}},\ \bibinfo {pages} {2117} (\bibinfo {year} {1994})}\BibitemShut
  {NoStop}%
\bibitem [{\citenamefont {Le}\ \emph {et~al.}(2016)\citenamefont {Le},
  \citenamefont {Wei}, \citenamefont {Jin},\ and\ \citenamefont
  {Lin}}]{Le2016}%
  \BibitemOpen
  \bibfield  {author} {\bibinfo {author} {\bibfnamefont {A.-T.}\ \bibnamefont
  {Le}}, \bibinfo {author} {\bibfnamefont {H.}~\bibnamefont {Wei}}, \bibinfo
  {author} {\bibfnamefont {C.}~\bibnamefont {Jin}},\ and\ \bibinfo {author}
  {\bibfnamefont {C.~D.}\ \bibnamefont {Lin}},\ }\bibfield  {title} {\bibinfo
  {title} {Strong-field approximation and its extension for high-order harmonic
  generation with mid-infrared lasers},\ }\href
  {https://doi.org/10.1088/0953-4075/49/5/053001} {\bibfield  {journal}
  {\bibinfo  {journal} {Journal of Physics B: Atomic, Molecular and Optical
  Physics}\ }\textbf {\bibinfo {volume} {49}},\ \bibinfo {pages} {053001}
  (\bibinfo {year} {2016})}\BibitemShut {NoStop}%
\bibitem [{\citenamefont {Milo\ifmmode \check{s}\else
  \v{s}\fi{}evi\ifmmode~\acute{c}\else \'{c}\fi{}}\ and\ \citenamefont
  {Becker}(2002)}]{Milo2002}%
  \BibitemOpen
  \bibfield  {author} {\bibinfo {author} {\bibfnamefont {D.~B.}\ \bibnamefont
  {Milo\ifmmode \check{s}\else \v{s}\fi{}evi\ifmmode~\acute{c}\else
  \'{c}\fi{}}}\ and\ \bibinfo {author} {\bibfnamefont {W.}~\bibnamefont
  {Becker}},\ }\bibfield  {title} {\bibinfo {title} {Role of long quantum
  orbits in high-order harmonic generation},\ }\href
  {https://doi.org/10.1103/PhysRevA.66.063417} {\bibfield  {journal} {\bibinfo
  {journal} {Phys. Rev. A}\ }\textbf {\bibinfo {volume} {66}},\ \bibinfo
  {pages} {063417} (\bibinfo {year} {2002})}\BibitemShut {NoStop}%
\bibitem [{\citenamefont {Sansone}\ \emph {et~al.}(2004)\citenamefont
  {Sansone}, \citenamefont {Vozzi}, \citenamefont {Stagira},\ and\
  \citenamefont {Nisoli}}]{Sansone2004}%
  \BibitemOpen
  \bibfield  {author} {\bibinfo {author} {\bibfnamefont {G.}~\bibnamefont
  {Sansone}}, \bibinfo {author} {\bibfnamefont {C.}~\bibnamefont {Vozzi}},
  \bibinfo {author} {\bibfnamefont {S.}~\bibnamefont {Stagira}},\ and\ \bibinfo
  {author} {\bibfnamefont {M.}~\bibnamefont {Nisoli}},\ }\bibfield  {title}
  {\bibinfo {title} {Nonadiabatic quantum path analysis of high-order harmonic
  generation: Role of the carrier-envelope phase on short and long paths},\
  }\href {https://doi.org/10.1103/PhysRevA.70.013411} {\bibfield  {journal}
  {\bibinfo  {journal} {Phys. Rev. A}\ }\textbf {\bibinfo {volume} {70}},\
  \bibinfo {pages} {013411} (\bibinfo {year} {2004})}\BibitemShut {NoStop}%
\bibitem [{\citenamefont {Keldysh}(1965)}]{Keldysh1965}%
  \BibitemOpen
  \bibfield  {author} {\bibinfo {author} {\bibfnamefont {L.~V.}\ \bibnamefont
  {Keldysh}},\ }\bibfield  {title} {\bibinfo {title} {{Ionization in the Field
  of a Strong Electromagnetic Wave}},\ }\href@noop {} {\bibfield  {journal}
  {\bibinfo  {journal} {J. Exp. Theor. Phys.}\ }\textbf {\bibinfo {volume}
  {20}},\ \bibinfo {pages} {1307} (\bibinfo {year} {1965})}\BibitemShut
  {NoStop}%
\bibitem [{\citenamefont {Milo\ifmmode \check{s}\else
  \v{s}\fi{}evi\ifmmode~\acute{c}\else \'{c}\fi{}}\ \emph
  {et~al.}(2000)\citenamefont {Milo\ifmmode \check{s}\else
  \v{s}\fi{}evi\ifmmode~\acute{c}\else \'{c}\fi{}}, \citenamefont {Becker},\
  and\ \citenamefont {Kopold}}]{Milo2000a}%
  \BibitemOpen
  \bibfield  {author} {\bibinfo {author} {\bibfnamefont {D.~B.}\ \bibnamefont
  {Milo\ifmmode \check{s}\else \v{s}\fi{}evi\ifmmode~\acute{c}\else
  \'{c}\fi{}}}, \bibinfo {author} {\bibfnamefont {W.}~\bibnamefont {Becker}},\
  and\ \bibinfo {author} {\bibfnamefont {R.}~\bibnamefont {Kopold}},\
  }\bibfield  {title} {\bibinfo {title} {Generation of circularly polarized
  high-order harmonics by two-color coplanar field mixing},\ }\href
  {https://doi.org/10.1103/PhysRevA.61.063403} {\bibfield  {journal} {\bibinfo
  {journal} {Phys. Rev. A}\ }\textbf {\bibinfo {volume} {61}},\ \bibinfo
  {pages} {063403} (\bibinfo {year} {2000})}\BibitemShut {NoStop}%
\bibitem [{\citenamefont {Milo\ifmmode \check{s}\else
  \v{s}\fi{}evi\ifmmode~\acute{c}\else \'{c}\fi{}}\ and\ \citenamefont
  {Becker}(2000{\natexlab{a}})}]{Milo2000b}%
  \BibitemOpen
  \bibfield  {author} {\bibinfo {author} {\bibfnamefont {D.~B.}\ \bibnamefont
  {Milo\ifmmode \check{s}\else \v{s}\fi{}evi\ifmmode~\acute{c}\else
  \'{c}\fi{}}}\ and\ \bibinfo {author} {\bibfnamefont {W.}~\bibnamefont
  {Becker}},\ }\bibfield  {title} {\bibinfo {title} {Attosecond pulse trains
  with unusual nonlinear polarization},\ }\href
  {https://doi.org/10.1103/PhysRevA.62.011403} {\bibfield  {journal} {\bibinfo
  {journal} {Phys. Rev. A}\ }\textbf {\bibinfo {volume} {62}},\ \bibinfo
  {pages} {011403} (\bibinfo {year} {2000}{\natexlab{a}})}\BibitemShut
  {NoStop}%
\bibitem [{\citenamefont {Pisanty}\ \emph {et~al.}(2014)\citenamefont
  {Pisanty}, \citenamefont {Sukiasyan},\ and\ \citenamefont
  {Ivanov}}]{Pisanty2014}%
  \BibitemOpen
  \bibfield  {author} {\bibinfo {author} {\bibfnamefont {E.}~\bibnamefont
  {Pisanty}}, \bibinfo {author} {\bibfnamefont {S.}~\bibnamefont {Sukiasyan}},\
  and\ \bibinfo {author} {\bibfnamefont {M.}~\bibnamefont {Ivanov}},\
  }\bibfield  {title} {\bibinfo {title} {Spin conservation in
  high-order-harmonic generation using bicircular fields},\ }\href
  {https://doi.org/10.1103/PhysRevA.90.043829} {\bibfield  {journal} {\bibinfo
  {journal} {Phys. Rev. A}\ }\textbf {\bibinfo {volume} {90}},\ \bibinfo
  {pages} {043829} (\bibinfo {year} {2014})}\BibitemShut {NoStop}%
\bibitem [{\citenamefont {Kosmann-Schwarzbach}(2011)}]{Noether2011}%
  \BibitemOpen
  \bibfield  {author} {\bibinfo {author} {\bibfnamefont {Y.}~\bibnamefont
  {Kosmann-Schwarzbach}},\ }\bibfield  {title} {\bibinfo {title} {Emmy
  noether’s wonderful theorem},\ }\href {https://doi.org/10.1063/PT.3.1263}
  {\bibfield  {journal} {\bibinfo  {journal} {Physics Today}\ }\textbf
  {\bibinfo {volume} {64}},\ \bibinfo {pages} {62} (\bibinfo {year} {2011})},\
  \Eprint {https://arxiv.org/abs/https://doi.org/10.1063/PT.3.1263}
  {https://doi.org/10.1063/PT.3.1263} \BibitemShut {NoStop}%
\bibitem [{\citenamefont {Bandres}\ and\ \citenamefont
  {Guizar-Sicairos}(2009)}]{Bandres2009}%
  \BibitemOpen
  \bibfield  {author} {\bibinfo {author} {\bibfnamefont {M.~A.}\ \bibnamefont
  {Bandres}}\ and\ \bibinfo {author} {\bibfnamefont {M.}~\bibnamefont
  {Guizar-Sicairos}},\ }\bibfield  {title} {\bibinfo {title} {Paraxial group},\
  }\href {https://doi.org/10.1364/OL.34.000013} {\bibfield  {journal} {\bibinfo
   {journal} {Opt. Lett.}\ }\textbf {\bibinfo {volume} {34}},\ \bibinfo {pages}
  {13} (\bibinfo {year} {2009})}\BibitemShut {NoStop}%
\bibitem [{\citenamefont {Adams}(2004)}]{Adams2004}%
  \BibitemOpen
  \bibfield  {author} {\bibinfo {author} {\bibfnamefont {C.}~\bibnamefont
  {Adams}},\ }\href {https://books.google.de/books?id=RqiMCgAAQBAJ} {\emph
  {\bibinfo {title} {The Knot Book: An Elementary Introduction to the
  Mathematical Theory of Knots}}}\ (\bibinfo  {publisher} {American
  Mathematical Society},\ \bibinfo {year} {2004})\BibitemShut {NoStop}%
\bibitem [{\citenamefont {Milošević}(2019)}]{Milo2019}%
  \BibitemOpen
  \bibfield  {author} {\bibinfo {author} {\bibfnamefont {D.~B.}\ \bibnamefont
  {Milošević}},\ }\bibfield  {title} {\bibinfo {title} {Quantum-orbit
  analysis of high-order harmonic generation by bicircular field},\ }\href
  {https://doi.org/10.1080/09500340.2018.1511862} {\bibfield  {journal}
  {\bibinfo  {journal} {Journal of Modern Optics}\ }\textbf {\bibinfo {volume}
  {66}},\ \bibinfo {pages} {47} (\bibinfo {year} {2019})},\ \Eprint
  {https://arxiv.org/abs/https://doi.org/10.1080/09500340.2018.1511862}
  {https://doi.org/10.1080/09500340.2018.1511862} \BibitemShut {NoStop}%
\bibitem [{\citenamefont {Hern\'andez-Garc\'{\i}a}\ \emph
  {et~al.}(2013)\citenamefont {Hern\'andez-Garc\'{\i}a}, \citenamefont
  {Pic\'on}, \citenamefont {San~Rom\'an},\ and\ \citenamefont
  {Plaja}}]{Garcia2013}%
  \BibitemOpen
  \bibfield  {author} {\bibinfo {author} {\bibfnamefont {C.}~\bibnamefont
  {Hern\'andez-Garc\'{\i}a}}, \bibinfo {author} {\bibfnamefont
  {A.}~\bibnamefont {Pic\'on}}, \bibinfo {author} {\bibfnamefont
  {J.}~\bibnamefont {San~Rom\'an}},\ and\ \bibinfo {author} {\bibfnamefont
  {L.}~\bibnamefont {Plaja}},\ }\bibfield  {title} {\bibinfo {title}
  {Attosecond extreme ultraviolet vortices from high-order harmonic
  generation},\ }\href {https://doi.org/10.1103/PhysRevLett.111.083602}
  {\bibfield  {journal} {\bibinfo  {journal} {Phys. Rev. Lett.}\ }\textbf
  {\bibinfo {volume} {111}},\ \bibinfo {pages} {083602} (\bibinfo {year}
  {2013})}\BibitemShut {NoStop}%
\bibitem [{\citenamefont {Milo\ifmmode \check{s}\else
  \v{s}\fi{}evi\ifmmode~\acute{c}\else \'{c}\fi{}}\ and\ \citenamefont
  {Becker}(2000{\natexlab{b}})}]{PhysRevA.62.011403}%
  \BibitemOpen
  \bibfield  {author} {\bibinfo {author} {\bibfnamefont {D.~B.}\ \bibnamefont
  {Milo\ifmmode \check{s}\else \v{s}\fi{}evi\ifmmode~\acute{c}\else
  \'{c}\fi{}}}\ and\ \bibinfo {author} {\bibfnamefont {W.}~\bibnamefont
  {Becker}},\ }\bibfield  {title} {\bibinfo {title} {Attosecond pulse trains
  with unusual nonlinear polarization},\ }\href
  {https://doi.org/10.1103/PhysRevA.62.011403} {\bibfield  {journal} {\bibinfo
  {journal} {Phys. Rev. A}\ }\textbf {\bibinfo {volume} {62}},\ \bibinfo
  {pages} {011403} (\bibinfo {year} {2000}{\natexlab{b}})}\BibitemShut
  {NoStop}%
\end{thebibliography}%
\newpage
\end{document}